%% file: ANA-STDM-2018-49-PAPER.tex
\pdfoutput=1
\documentclass[cernpreprint, atlasdraft=false, texlive=2023, UKenglish, texmf, orcidlogo]{atlasdoc}
 
\usepackage{atlaspackage}
\usepackage{atlasbiblatex}
 
\usepackage{atlasphysics}
 
\graphicspath{{logos/}{figures/}}
 
\usepackage{ANA-STDM-2018-49-PAPER-defs}

\AtlasTitle{Cross-section measurements for the production of a $Z$ boson in association with high-transverse-momentum jets in $pp$ collisions at $\sqrt{s} = 13$~TeV with the ATLAS detector}
 
\AtlasAbstract{
Cross-section measurements for a $Z$ boson produced in association with high-transverse-momentum jets ($p_{\mathrm{T}} \geq 100$~GeV) and decaying into a charged-lepton pair ($e^+e^-,\mu^+\mu^-$) are presented. The measurements are performed using proton--proton collisions at $\sqrt{s}=13$~TeV corresponding to an integrated luminosity of $139$~fb$^{-1}$ collected by the ATLAS experiment at the LHC. Measurements of angular correlations between the $Z$ boson and the closest jet are performed in events with at least one jet with $p_{\mathrm{T}} \geq 500$~GeV. Event topologies of particular interest are the collinear emission of a $Z$ boson in dijet events and a boosted $Z$ boson recoiling against a jet. Fiducial cross sections are compared with state-of-the-art theoretical predictions. The data are found to agree with next-to-next-to-leading-order predictions by \textsc{NNLOjet} and with the next-to-leading-order multi-leg  generators \textsc{MadGraph5\_aMC@NLO} and \textsc{Sherpa}.
}

\author{The ATLAS Collaboration}
 
\AtlasRefCode{STDM-2018-49}
 
\AtlasDate{\today}
 
\AtlasJournal{JHEP}
\AtlasJournalRef{JHEP 06 (2023) 080}
\AtlasDOI{10.1007/JHEP06(2023)080}
 
\AtlasHeading{ATLAS}

\PreprintIdNumber{CERN-EP-2022-075}

\hypersetup{pdftitle={ATLAS document},pdfauthor={The ATLAS Collaboration}}
 
\begin{document}
 
\maketitle

\section{Introduction}
\label{sec:Introduction}
 
The measurement of \Zdash-boson production\footnote{Throughout this paper, \Zg-boson production is simply referred to as \Zdash-boson production.}  in association with jets, \Zjets, constitutes a powerful test of perturbative quantum chromodynamics (QCD)~\cite{Gross:1973, Politzer:1974} and, in the case of high-energy jets, it provides a way to probe the interplay between QCD and higher-order electroweak (EW) processes~\cite{Rubin:2010,Christiansen:2016,Boughezal:2015,Christiansen:2014}.
The large \Zjets\ production cross section and the easily identifiable decays of the \ZBoson\ to charged-lepton final states offer a clean experimental signature which can be measured precisely. Such processes also constitute non-negligible backgrounds in measurements of the Higgs boson~\cite{HIGG-2018-52,HIGG-2017-14} and in searches for new phenomena~\cite{SUSY-2018-10,SUSY-2018-16,CMS-B2G-19-006}, which often exploit the presence of high-\pt\ jets to enrich a data sample with potential signal. In those studies, predictions are used to extrapolate \Zjets\ backgrounds from control regions to the signal regions and to model the distributions of the final discriminants.
 
In the calculations of \Zonejet\ production at leading order (LO), the \ZBoson\ recoils against a quark or a gluon. At next-to-leading order (NLO), real and virtual QCD and EW effects play a role in  \Zjets\ production, such as in topologies corresponding to dijet events where a real \ZBoson\ is emitted from an incoming or outgoing quark leg~\cite{Rubin:2010,Christiansen:2016,Boughezal:2015,Christiansen:2014}. Example Feynman diagrams for LO and NLO \Zjets\ production processes are shown in Figure~\ref{fig:feynmans}. The latter case can lead to production rate enhancements proportional to $\alphas \ln^2(\ptlj/\mZ)$, where $\alphas$ is the strong coupling constant, $\ptlj$ the transverse momentum of the leading jet, and $\mZ$ the mass of the \ZBoson, and thus the effect can become very large for events with high-\pT jets.
These events exhibit a collinear enhancement in the distribution of the angular distance between the \ZBoson\ and the closest jet. Although the enhancement can be probed in the region of small angular separation, this region also contains contributions where the \ZBoson\ is produced in association with larger numbers of jets, which must be included in the predictions. The measurements presented in this paper target QCD-only \Zjets\ production, treating EW \Ztwojets\ (EW $Zjj$) production~\cite{STDM-2017-27} as a background. Measurements where the EW $Zjj$ contribution is treated as signal and not subtracted as background are also performed and published in the HEPData entry~\cite{hepdata} of this measurement.

The ATLAS Collaboration~\cite{PERF-2007-01} at the Large Hadron Collider (LHC)~\cite{Evans:2008zzb} first measured angular distributions in high-\pt\ \WBoson\ production with jets (\Wjets) in the 8~\TeV\ $pp$-collision data set~\cite{STDM-2015-16}.
The first similar measurement in \Zjets\ events was published by the CMS Collaboration and used a partial 13~\TeV\ data set corresponding to $35.9~\ifb$~\cite{CMS-SMP-19-010}. Both measurements highlight the fact that the collinear region, where the angular separation between the \WZBoson\ and the closest jet is small, represents a major challenge for contemporary Monte Carlo (MC) generators. The measurements presented in this paper include a wide range of new observables sensitive to the presence of high-\pT jets and to the collinear emission of a \ZBoson\ in dijet events. The statistical power of the full LHC Run-2 data set makes it possible to tighten the collinear selection, and to measure key observables separately for collinear and non-collinear topologies.

This publication focuses on events that contain a \Zdash-boson candidate reconstructed from either an $e^{+}e^{-}$ or $\mu^{+}\mu^{-}$ pair in association with hadronic jets defined as jets having transverse momentum greater than or equal to $100~\GeV$.
The phase-space region with at least one associated jet is labelled as the \inc\ region. In this region, the measured quantities are the transverse momentum of the leading jet (\ptlj), the transverse momentum of the \ZBoson\ (\ptZ), the scalar sum of the transverse momentum of all selected jets and leptons (\nht), and the jet multiplicity.
A \hpt\ region is selected by requiring the presence of a jet with $\pT \geq 500~\GeV$. To test the prediction that this region is composed of two characteristic topologies, the soft radiation of a \ZBoson\ from a jet (\col\ topology) and the hard scatter of a \ZBoson\ against a jet (\btb\ topology), the \hpt\ region is split to cover different ranges of the angle between the \ZBoson\ and the closest jet, and selected key observables are measured separately for each region.
In the \hpt\ region, the following observables sensitive to the presence of collinear \Zdash-boson emission are studied:
\begin{itemize}
\item \mDRZj, the angular distance\footnote{\AtlasCoordFootnote} between the \ZBoson\ and the closest jet. Real \Zdash-boson radiation is expected to be enhanced at low values of \mDRZj. At large values of \mDRZj, the \ZBoson\ is balanced by a recoiling jet and large virtual EW corrections are expected. To enrich these two topologies, \col\ and \btb\ regions are constructed by requiring $\mDRZj \leq 1.4$ and $\mDRZj \geq 2.0$, respectively.
\item \ptzj, the ratio of the \Zdash-boson \pT\ to the closest-jet \pT, defined as
\begin{equation*}
\ptzj \equiv \frac{\ptZ}{\ptcj}.
\end{equation*}
Collinear \Zdash-boson radiation is expected to be dominated by soft \ZBosons, resulting in very small values for this ratio.
\item $N_\mathrm{jets}$, the jet multiplicity. The \btb\ region is expected to be dominated by \Zonejet\ events, whereas the \col\ region  would be dominated by \Ztwojets\ events.
\end{itemize}
The measurements of jet multiplicity and \ptzj\ are performed both in the full \hpt\ region and separately in the \col\ and \btb\ regions.
 
\begin{figure*}[t]
\centering
\subfloat{
\includegraphics[width=.35\textwidth]{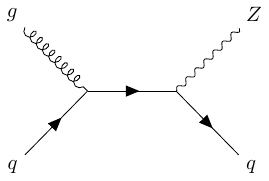}
}
\subfloat{
\includegraphics[width=.35\textwidth]{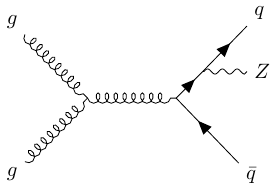}
}
\caption{
Representative Feynman diagrams for the production of a \ZBoson\ in association with high-\pT\ jets. The \Zonejet\ events (left) are expected to populate the \btb\ region where the \ZBoson\ is balanced against a single high-\pT\ jet. In dijet events (right), the \ZBoson\ is expected to be radiated from the quark leg, with kinematics leading to small values of the angular distance between the \ZBoson\ and the closest jet, \minDR , and therefore populating the \col\ region.
}
\label{fig:feynmans}
\end{figure*}
 
Measurements of jet multiplicity and \mDRZj\ are also performed in an alternative high-energy region, constructed by requiring the scalar sum of the transverse momentum of all selected jets, \htj, to be at least $600~\GeV$. This alternative region probes high-energy events but does not depend on the presence of a single very energetic object. In this region, called the \hht\ region, a large fraction of the events have higher jet multiplicity.
 
Predictions from the most recent generators combine NLO multi-leg matrix elements (ME) with a parton shower (PS) and hadronisation model~\cite{Bothmann:2019yzt,PMGR-2021-01,Frederix_2012,Sjostrand:2014zea}. Fixed-order parton-level theoretical predictions for \Zjets\ production at next-to-next-to-leading order (NNLO) are available for up to one associated jet~\cite{Boughezal:2016a,Boughezal:2016b,Gehrman:2016a,Gehrman:2016b}. In this paper, the cross-section measurements are compared with state-of-the-art multi-leg ME+PS generators and NNLO fixed-order \Zjets\ predictions from \NNLOJET~\cite{Gehrman:2016a,Gehrman:2016b}. Virtual EW corrections were made available recently~\cite{Kallweit:2015dum,Kallweit:2014xda} and are included in one of the \Sherpa~\cite{PMGR-2021-01} predictions studied in this paper.
 
The paper is organised as follows. Section~\ref{sec:ATLASDet} contains a brief overview  of the ATLAS detector. The data and simulated samples, as well as additional predictions used in the analysis, are described in Section~\ref{sec:DataMCSamples}. The object definition and the event reconstruction at detector level are presented in Section~\ref{sec:Reconstruction}, while Section~\ref{sec:Backgrounds} describes the background modelling and presents a comparison of measured and predicted yields at detector level. After background subtraction, the data are unfolded to particle level in a fiducial phase space with a procedure described in Section~\ref{sec:Unfolding}. The experimental and theoretical systematic uncertainties are estimated in Section~\ref{sec:Systematics}. Section~\ref{sec:Results} presents the unfolded cross-section results and the comparisons with predictions. Conclusions are provided in Section~\ref{sec:Conclusions}.


\section{ATLAS detector}
\label{sec:ATLASDet}
 
The ATLAS experiment at the LHC is a multipurpose particle detector
with a forward--backward symmetric cylindrical geometry and a near \(4\pi\) coverage in
solid angle.
It consists of an inner tracking detector surrounded by a thin superconducting solenoid
providing a \SI{2}{\tesla} axial magnetic field, electromagnetic and hadron calorimeters, and a muon spectrometer.
The inner tracking detector covers the pseudorapidity range \(|\eta| < 2.5\).
It consists of silicon pixel, silicon microstrip, and transition radiation tracking detectors.
Lead/liquid-argon (LAr) sampling calorimeters provide electromagnetic (EM) energy measurements
with high granularity.
A steel/scintillator-tile hadron calorimeter covers the central pseudorapidity range (\(|\eta| < 1.7\)).
The endcap and forward regions are instrumented with LAr calorimeters
for both the EM and hadronic energy measurements up to \(|\eta| = 4.9\).
The muon spectrometer surrounds the calorimeters and is based on
three large superconducting air-core toroidal magnets with eight coils each.
The field integral of the toroids ranges between \num{2.0} and \SI{6.0}{\tesla\metre}
across most of the detector.
The muon spectrometer includes a system of precision tracking chambers and fast detectors for triggering.
A two-level trigger system is used to select events.
The first-level trigger is implemented in hardware and uses a subset of the detector information
to accept events at a rate below \SI{100}{\kHz}.
This is followed by a software-based trigger that
reduces the accepted event rate to \SI{1}{\kHz} on average
depending on the data-taking conditions.
An extensive software suite~\cite{ATL-SOFT-PUB-2021-001} is used in the reconstruction and analysis of real
and simulated data, in detector operations, and in the trigger and data acquisition systems of the experiment.


\section{Data set and simulated event samples}
\label{sec:DataMCSamples}
 
The data used in this analysis were recorded with the ATLAS detector from 2015 to 2018 in $pp$ collisions at $\sqrt{s}=$ 13~\TeV\ (full Run-2 data set) and correspond to a total integrated luminosity of 139~\ifb~\cite{DAPR-2011-01}. The mean number of $pp$ interactions per bunch crossing, including the hard scattering and other interactions in the same and neighbouring bunch crossings (pile-up), was $\langle \mu \rangle = 34$.
 
MC simulation samples are used to estimate most of the contributions from background events, to unfold the data to particle level, and in comparisons with the unfolded data distributions. The generated samples were processed using the \GEANT-based ATLAS detector simulation~\cite{SOFT-2010-01,Agostinelli:2002hh} and the same event-reconstruction algorithms are used for both the MC samples and the data. A summary of the MC generators and calculations used for the simulation of signal and background processes is provided in Table~\ref{tab:generators}.

 
\begin{table}
\centering
\renewcommand{\arraystretch}{1.1}
\caption{Summary of the programs used to produce the signal and the various background samples. For every process the name of the program used is indicated in the second column. The third column reports the order of the QCD calculation in the matrix elements, where $n$p denotes the number of real parton emissions. The \Shtte\ \Zjets\ processes include virtual electroweak corrections.
}
\begin{tabular}{c | c c c }
\toprule
\multicolumn{1}{c}{Process}  & Generator  & Order pQCD  & References  \\
\midrule
\multicolumn{1}{c}{Signal} \\
\midrule
$Z \to \ell\ell$ ($\ell=e, \mu)$ & \Shtte\ & 0--2p NLO, 3--5p LO & \cite{PMGR-2021-01,Gleisberg:2008fv,Buccioni:2019sur,Cascioli:2011va,Denner:2016kdg,Schumann:2007mg,Hoeche:2011fd,Hoeche:2012yf,Catani:2001cc,Hoeche:2009rj,Catani:2002hc,Ball:2014uwa} \\
$Z \to \ell\ell$ ($\ell=e, \mu$) & \MGfnlo\ 	& 0--3p NLO & \cite{Frederix_2012,Sjostrand:2014zea,Alwall:2014hca,ATL-PHYS-PUB-2014-021,Ball:2012cx,Ball:2014uwa,PMGR-2021-01} \\
\hline
$Z \to \ell\ell$ ($\ell=e, \mu$)	& \Shtto\ & 0--2p NLO, 3--4p LO & \cite{Bothmann:2019yzt,Gleisberg:2008fv,Buccioni:2019sur,Cascioli:2011va,Denner:2016kdg,Schumann:2007mg,Hoeche:2011fd,Hoeche:2012yf,Catani:2001cc,Hoeche:2009rj}\\
$Z \to \ell\ell$ ($\ell=e, \mu$) & \MGflo\ & 0--4p LO & \cite{Sjostrand:2007gs,Lonnblad:2001iq,Lonnblad:2011xx,Alwall:2014hca} \\
\hline
$Z \to \ell\ell$ ($\ell=e, \mu$) & \njnnlo\ & 1p NNLO & \cite{Gehrman:2016a,Gehrman:2016b} \\
$Z \to \ell\ell$ ($\ell=e, \mu$) & \njnlo\ & 1p NLO & \cite{Gehrman:2016a,Gehrman:2016b} \\
\midrule
\multicolumn{1}{c}{Backgrounds} \\
\midrule
EW $Zjj$($\to$ $\ell\ell$ ($\ell=e, \mu$))  & \HERWIG[7.1.5], \VBFNLO[3.0.0]  & NLO & ~\cite{Bahr:2008pv,Bellm:2015jjp,Arnold:2011wj}\\
\hline
$Z\to \tau\tau$	& \Shtto\ & 0--2p NLO, 3--4p LO & \cite{Bothmann:2019yzt,Gleisberg:2008fv,Buccioni:2019sur,Cascioli:2011va,Denner:2016kdg,Schumann:2007mg,Hoeche:2011fd,Hoeche:2012yf,Catani:2001cc,Hoeche:2009rj}\\
\Wjets	& \Shtto\ & 0--2p NLO, 3--4p LO  & \cite{Bothmann:2019yzt,Gleisberg:2008fv,Buccioni:2019sur,Cascioli:2011va,Denner:2016kdg,Schumann:2007mg,Hoeche:2011fd,Hoeche:2012yf,Catani:2001cc,Hoeche:2009rj} \\
\hline
$t\bar{t}$ & \POWHEGBOX[v2] + \PYTHIA[8.230]   & NLO & \cite{Frixione:2007nw,Nason:2004rx,Frixione:2007vw,Alioli:2010xd} \\
\hline
Single top ($t$-, $Wt$-, $s$-channel)   & \POWHEGBOX[v2] + \PYTHIA[8.230]   & NLO & \cite{Frixione:2007nw,Nason:2004rx,Frixione:2007vw,Alioli:2010xd} \\
\hline
$Z/W$($\to qq$)$Z$($\to  \ell\ell$) &  \Shtto\ & 0--1p NLO, 2--3p LO & \cite{Bothmann:2019yzt,Gleisberg:2008fv,Buccioni:2019sur,Cascioli:2011va,Denner:2016kdg,Schumann:2007mg,Hoeche:2011fd,Hoeche:2012yf,Catani:2001cc,Hoeche:2009rj} \\
$W$($\to \ell \nu$)$Z$($\to qq$)	& \Shtto\ & 0--1p NLO, 2--3p LO & \cite{Bothmann:2019yzt,Gleisberg:2008fv,Buccioni:2019sur,Cascioli:2011va,Denner:2016kdg,Schumann:2007mg,Hoeche:2011fd,Hoeche:2012yf,Catani:2001cc,Hoeche:2009rj} \\
$W^{\pm}$($\to qq$)$W^{\mp}$($\to \ell \nu$)	& \Shtto\ & 0--1p NLO, 2--3p LO & \cite{Bothmann:2019yzt,Gleisberg:2008fv,Buccioni:2019sur,Cascioli:2011va,Denner:2016kdg,Schumann:2007mg,Hoeche:2011fd,Hoeche:2012yf,Catani:2001cc,Hoeche:2009rj} \\
$\ell\ell\nu\nu, \ell\ell\ell\nu, \ell\ell\ell\ell$ & \Shttt\ & 0--1p NLO, 2--3p LO & \cite{Bothmann:2019yzt,Gleisberg:2008fv,Buccioni:2019sur,Cascioli:2011va,Denner:2016kdg,Schumann:2007mg,Hoeche:2011fd,Hoeche:2012yf,Catani:2001cc,Hoeche:2009rj} \\
\hline
$V$($\to \ell\ell$)\,+\,$\gamma$ & \Shtteight\ & 0--1p NLO, 2--3p LO & \cite{Bothmann:2019yzt,Gleisberg:2008fv,Buccioni:2019sur,Cascioli:2011va,Denner:2016kdg,Schumann:2007mg,Hoeche:2011fd,Hoeche:2012yf,Catani:2001cc,Hoeche:2009rj} \\
 
\bottomrule
\end{tabular}
\label{tab:generators}
\end{table}


The production of \ZBosons\ in association with jets was simulated with the ATLAS configuration of \Shtte~\cite{PMGR-2021-01}, which  includes matrix elements for up to five partons at LO and up to two partons at NLO. They are calculated with the Comix~\cite{Gleisberg:2008fv} and \OpenLoops~\cite{Buccioni:2019sur,Cascioli:2011va,Denner:2016kdg} libraries and  matched with the \SHERPA parton shower~\cite{Schumann:2007mg} using the \MEandPSatNLO prescription~\cite{Hoeche:2011fd,Hoeche:2012yf,Catani:2001cc,Hoeche:2009rj} with a set of tuned parameters (`tune') developed by the \SHERPA authors.
In contrast to \Shtto~\cite{Bothmann:2019yzt}, used previously in ATLAS publications, it includes a modified Catani--Seymour subtraction scheme~\cite{Catani:2002hc}, the Hessian \NNPDFthreeNNLO\ PDF set~\cite{Ball:2014uwa} is used, and an analytic enhancement technique has been introduced~\cite{PMGR-2021-01}.
The cross section in the \hpt\ region has been considerably reduced relative to the prediction from the previous \SHERPA version by switching to an improved matching scheme with a different treatment of unordered histories~\cite{Hoeche:2009rj}. \Shtte\ is also the only sample used in this paper which includes NLO virtual EW corrections~\cite{Cascioli:2011va,Denner:2016kdg}. The samples were produced using three options for the combination of NLO EW and QCD corrections: an additive, a multiplicative, and an exponential scheme. The nominal prediction is derived via the additive scheme; a systematic uncertainty band is derived from the envelope of all schemes. In contrast to virtual EW corrections, EW parton showers are not included in any of the generators used in this paper. The \Shtte\ \Zjets\ samples are used for the nominal unfolding of the data distributions, to estimate the systematic uncertainties and in comparisons with the cross-section measurements.
 
A second \Zjets\ sample, referred to as \MGfnlo~\cite{PMGR-2021-01}, was produced by using the \MGNLO[2.6.5]~\cite{Alwall:2014hca} program to generate matrix elements at NLO accuracy in QCD for up to three additional partons in the final state. The \NNPDFthreeOneNNLO\ set~\cite{Ball:2014uwa} was used in the generation. The parton showering and subsequent hadronisation was performed using \PYTHIA[8.240]~\cite{Sjostrand:2014zea} with the A14 tune~\cite{ATL-PHYS-PUB-2014-021} and the \NNPDFtwoLO PDF set~\cite{Ball:2012cx}. The jet multiplicities were merged using the FxFx prescription~\cite{Frederix_2012}. This prediction is compared with the unfolded cross-section measurements.

A third sample of \Zjets\ events and an event sample from \Wjets\ processes were produced with the \Shtto~\cite{Bothmann:2019yzt} generator using NLO matrix elements for up to two partons, and LO matrix elements for up to four partons, calculated with the Comix and \OpenLoops\ libraries. They were matched with the \SHERPA parton shower using the \MEandPSatNLO prescription with a set of tuned parameters developed by the \SHERPA authors. The MC replica version of the \NNPDFthreeNNLO\ set of PDFs was used. The \Shtto\ \Zjets\ sample is used in comparisons with the unfolded cross-section measurements, as it was used as a standard in previous ATLAS Run-2 publications.
 
A fourth \Zjets\ sample, referred to as \MGflo, was generated using LO-accurate matrix elements with up to four final-state partons calculated by \MGNLO[2.2.2]~\cite{Alwall:2014hca}. The ME calculation employed the \NNPDFthreeNLO\ PDF set and was interfaced to \PYTHIA[8.186]~\cite{Sjostrand:2007gs} for the modelling of the parton shower, hadronisation, and underlying event. The overlap between matrix element and parton shower emissions was removed using the CKKW-L merging procedure~\cite{Lonnblad:2001iq,Lonnblad:2011xx}. The A14 tune of \PYTHIA was used with the \NNPDFtwoLO PDF set. This sample is used to validate the unfolding method and in comparisons with the unfolded cross-section measurements.
 
Two additional \Zjets\ samples were generated with the \NNLOJET\ program~\cite{Gehrman:2016a,Gehrman:2016b}, which computes fixed-order parton-level predictions for inclusive jet processes at higher orders in QCD. The NLO and NNLO predictions, referred to as \njnlo\ and \njnnlo, respectively, were calculated as higher-order corrections to the parton-level LO process of \Zonejet\ production.
The \NNPDFthreeOneNNLO\ set was used with a central scale choice of $\mu_0 = \frac{1}{2} (E_{\mathrm{T},Z} + \Sigma_{i\in \mathrm{partons}} p_{\mathrm{T},i})$ with $E_{\mathrm{T},Z} = \sqrt{m_{\ell\ell}^2 + p_{\mathrm{T},Z}^2}$. These samples are pure QCD predictions at parton level. To match the fiducial selection of the measurement (see Section~\ref{sec:Unfolding}), scale factors to correct from the Born level to the dressed-lepton level are computed and applied to these predictions. The slightly different overlap-removal procedure for jets and leptons used in these samples, due to the \NNLOJET\ program design, is addressed by overlap-removal correction scale factors. Both sets of scale factors, deviating from unity at the percent level and computed separately for each bin of the measured observables, are published in the HEPData entry~\cite{hepdata} of this measurement. Non-pertubative corrections are found to be consistent with zero when \ptlj\ exceeds 100~\GeV\ and are not needed to match the fiducial selection of these measurements. These samples are used in comparisons with the unfolded cross-section measurements.
 
The EW \Zjj\ process is defined by the $t$-channel exchange of a weak boson and at tree level is calculated at $\cal{O}$($\mathrm \alpha^4_\textsc{ew}$) when including the decay of the \Zboson\ boson~\cite{STDM-2017-27}. In contrast, the strong \Zjj\ process, which is covered by the \Zjets\ samples, has no weak boson exchanged in the $t$-channel and at tree level is calculated at $\cal{O}$($\mathrm \alpha^2_\textsc{ew}\alphas^2$) when including the decay of the \Zboson\ boson.
The EW \Zjj\ samples were produced in the vector-boson fusion (VBF) approximation with \HERWIG[7.1.5]~\cite{Bahr:2008pv,Bellm:2015jjp} at NLO accuracy in the strong coupling, using \VBFNLO[3.0.0]~\cite{Arnold:2011wj} to provide the loop amplitude. The \MMHT[lo] PDF set~\cite{Harland-Lang:2014zoa-CUSTOM} was used along with the default set of tuned parameters for parton showering, hadronisation and the underlying event. To account for the interference between strong \Zjj\ and EW \Zjj\ processes, a uniform modelling uncertainty of $25\%$ in the EW \Zjj\ cross section ($40\%$ in the collinear region), determined from simulation with \MGNLO[2.9.5], is applied~\cite{STDM-2017-27}.
 
The \ttbar\ background in this measurement is derived with a data-driven method as described in Section~\ref{sec:Backgrounds}. The MC \ttbar\ events used for intermediate steps of the method were modelled using the \POWHEGBOX[v2]~\cite{Frixione:2007nw,Nason:2004rx,Frixione:2007vw,Alioli:2010xd} generator at NLO with the \NNPDFthreeNLO\ PDF set and the \hdamp\ parameter\footnote{The \hdamp\ parameter is a resummation damping factor and one of the parameters that control the matching of \powheg matrix elements to the parton shower and thus effectively regulates the high-\pt\ radiation against which the \ttbar\ system recoils.} set to 1.5~\mtop~\cite{ATL-PHYS-PUB-2016-020}. The events were interfaced to \PYTHIA8.230~\cite{Sjostrand:2014zea} to model the parton shower, hadronisation, and underlying event, with parameters set according to the A14 tune  and using the \NNPDFtwoLO\ set of PDFs. The \ttbar\ sample is normalised to the cross-section prediction at \NNLO\ accuracy, including the resummation of next-to-next-to-leading logarithmic (NNLL) soft-gluon terms calculated with \TOPpp[2.0]~\cite{Beneke:2011mq,Cacciari:2011hy,Baernreuther:2012ws,Czakon:2012zr,Czakon:2012pz,Czakon:2013goa,Czakon:2011xx}.
 
Single top quark production in the $s$-channel, in the $t$-channel, and in association with a $W$ boson ($tW$) was modelled using the \POWHEGBOX[v2] generator at NLO in QCD with the five-flavour scheme and the \NNPDFthreeNLO\ set of PDFs. The diagram-removal scheme~\cite{Frixione:2008yi} was used to remove interference and overlap with \ttbar\ production. The $tW$ cross section is corrected to the theory prediction at approximate NNLO accuracy~\cite{Kidonakis:2010ux,Kidonakis:2013zqa}, while the $s$- and $t$-channel cross sections are corrected to the prediction at NLO accuracy~\cite{Kant:2014oha-CUSTOM,Aliev:2010zk}.
 
Samples of diboson final states ($VV$) were produced with the \Shtto\ or \Shttt\ generator depending on the process, including off-shell effects and Higgs boson contributions where appropriate. Fully leptonic final states and semileptonic final states, where one boson decays leptonically and the other hadronically, were generated using matrix elements at NLO accuracy in QCD for up to one additional parton and at LO accuracy for up to three additional parton emissions.
The matrix element calculations were matched and merged with the \SHERPA parton shower as detailed above for \Shtto, and the \NNPDFthreeNNLO set of PDFs was used.

The production of \Vgamma\ final states was simulated with the \Shtteight\ ~\cite{Bothmann:2019yzt} generator. Matrix elements at NLO QCD accuracy for up to one additional parton and LO accuracy for up to three additional parton emissions were matched and merged with the \SHERPA parton shower as detailed above for \Shtto, and the \NNPDFthreeNNLO set of PDFs was used.

Background events involving semileptonic decays of heavy quarks, hadrons misidentified as leptons, and, in the case of the electron channel, electrons from photon conversions are referred to collectively as `multijet events'. The multijet background is estimated using data-driven techniques, as described in Section~\ref{sec:Backgrounds}.
 
For bottom and charm hadron decays, the \EVTGEN[1.7.0] program~\cite{EvtGen} was used for \MGfnlo\ samples, and \EVTGEN[1.2.0] was used for all other \MADGRAPH and \POWHEG samples. The effect of multiple interactions in the same and neighbouring bunch crossings (\pileup) was modelled by overlaying the simulated hard-scattering event with inelastic $pp$ events generated with \PYTHIA[8.186]~\cite{Sjostrand:2007gs} using the \NNPDFtwoLO\ PDF and the A3 tune~\cite{ATL-PHYS-PUB-2016-017}. The small differences in lepton reconstruction, isolation, and trigger efficiencies between simulation and data are corrected in the simulation on an event-by-event basis by applying efficiency scale factors for each lepton~\cite{PERF-2018-01,PERF-2015-10,MUON-2018-03}.


\section{Event reconstruction}
\label{sec:Reconstruction}
 
Events are used if they were recorded during stable beam conditions and if they satisfy detector and data-quality requirements~\cite{DAPR-2018-01}. They are required to have a primary vertex, defined as the vertex with the highest sum of track $\pt^2$, with at least two associated tracks with $\pt > \SI{500}{\MeV}$~\cite{ATLAS:2010lca}. Events are selected using triggers~\cite{TRIG-2018-05,TRIG-2018-01,TRIG-2019-03-MANUAL} that require at least two electrons or two muons, or the combination of at least one electron and one muon; the efficiencies for these triggers plateau in the region of $\pt>\SI{25}{\GeV}$.
 
Electron candidates are reconstructed from inner-detector tracks which come from the primary vertex and are matched to clusters of energy deposits in the EM calorimeter.
To fulfil the primary-vertex condition, the electron track's transverse impact parameter significance must satisfy $|d_0|/\sigma(d_0) < 5.0$, where $d_0$ is the transverse impact parameter and $\sigma(d_0)$ its uncertainty, and the longitudinal impact parameter $z_0$ must satisfy $|z_0\sin(\theta)| < 0.5$~mm, where $\theta$ is the angle of the track to the beamline.
Electron candidates must satisfy the `Medium' likelihood-based identification requirements~\cite{PERF-2018-01} based on EM shower shapes, track quality, and track--cluster matching. They must also satisfy the `PflowLoose'~\cite{PERF-2018-01} isolation requirement. Electron candidates are used in the analysis if they have $\pt \geq 25~\GeV$ and $|\eta|<1.37$ or $1.52 < |\eta| < 2.47$.
 
Muon candidates are identified by matching inner-detector tracks from the primary vertex to either full tracks or track segments reconstructed in the muon spectrometer. The candidates must satisfy the following primary-vertex requirements: the transverse impact parameter significance must satisfy $|d_0|/\sigma(d_0) < 3.0$ and the longitudinal impact parameter must satisfy $|z_0\sin(\theta)| < 0.5$~mm, where $d_0$, $\sigma(d_0)$, $z_0$ and $\theta$ are as defined above for the electrons.  Muons are required to pass `Medium' identification requirements~\cite{PERF-2015-10,MUON-2018-03} based on quality criteria applied to the inner-detector and muon-spectrometer tracks. Muon candidates with $\pt \geq 300~\GeV$ must satisfy tighter identification requirements in the muon spectrometer in order to improve the muon-\pT resolution. Muons must also satisfy the `PflowLoose' isolation requirement, built from tracking and calorimeter information, with a muon-\pT-dependent variable cone size $\Delta R$~\cite{MUON-2018-03}. Muon candidates are used in the analysis if they have $\pt \geq 25~\GeV$ and $|\eta|<2.4$.
 
Jets of hadrons are reconstructed using a particle-flow algorithm~\cite{PERF-2015-09} based on noise-suppressed positive-energy topological clusters in the calorimeter. Energy deposited in the calorimeter by charged particles is subtracted and replaced by the momenta of tracks which are matched to those topological clusters. The jets are clustered using the anti-\kt~\cite{Cacciari:2008gp} algorithm implemented in the \textsc{FastJet} package~\cite{Cacciari:2011ma} with a radius parameter $R = 0.4$. They are further calibrated according to in situ measurements of the jet energy scale~\cite{JETM-2018-05}. Analysis jets are required to have a calibrated $\pT \geq 100~\GeV$ and $|y |<2.5$.
 
Electrons, muons and jets are reconstructed and identified independently. An overlap-removal procedure is then applied to uniquely identify these objects in an event. For the lepton--jet overlap removal, softer jets with $\pT \geq 30~\GeV$ and $| y |<2.5$ are considered. Preselected jets with a high probability to have been initiated by an electron or a radiated photon such that $\Delta R$ between the jet and a lepton is smaller than 0.2 are removed. In a second step, leptons closer than $\Delta R = 0.4$ to any remaining jet are removed.
 
Events are selected if they contain a \Zdash-boson candidate reconstructed from two same-flavour, opposite-charge leptons ($\ell = e, \mu$) and with dilepton invariant mass $71$~GeV $\leq \mll \leq 111$~GeV. The selected events are also required to contain at least one analysis jet. Events which satisfy the above selection requirements define the \inc\ \Zjets\ region. A dedicated \hpt\ region is created by requiring  the leading jet to have $\ptlj \geq 500~\GeV$. This latter region is split into the \col\ region, where the angular distance between the \ZBoson\ and the closest jet must be $\mDRZj \leq 1.4$, and the \btb\ region that requires $\mDRZj \geq 2.0$. An alternative phase space is also defined by requiring $\htj \geq 600~\GeV$, labelled as the \hht\ region, where \htj\ is defined as the scalar sum of the \pt\ of the analysis jets.


\section{Background estimation}
\label{sec:Backgrounds}
 
Backgrounds from single-boson, diboson and single-top-quark processes are estimated using the MC samples described in Section~\ref{sec:DataMCSamples}, while top-pair production and `multijet event' contributions from semileptonic decays of heavy quarks, hadrons misidentified as leptons, and electrons from photon conversions are estimated from data. A summary of the composition and relative importance of the backgrounds in the various signal regions is given in Table~\ref{tab:backgrounds}. The overall purity of the  \Zjets\ selections (defined as the expected fraction of signal events after the final selection) ranges from $94\%$ in the \inc\ region to $87\%$, $86\%$, $87\%$ and $88\%$ in the \hpt, \col, \btb\ and \hht\ regions, respectively. Backgrounds are dominated by \ttbar, diboson and EW \Zjj\ processes, with fractions of 2\%--5\% , 2\%--6\% and 1\%--5\%, respectively. The fraction of events arising from \Ztt, \Wjets, \Vgamma, and multijet backgrounds is below the percent level in all signal regions.

 
\def\arraystretch{1.1}
\begin{table}[h]
\caption{
Event yields in the different \Zjets\ signal regions in the electron and muon channels. Uncertainties correspond to the statistical and experimental systematic uncertainties added in quadrature. The \Zjets\ prediction is computed with \Shtte. The number of multijet events is negligible and not included.
}
\centering
\resizebox{\columnwidth}{!}{
\begin{tabular}{l| r @{$\pm$} l | r @{$\pm$} l |  r @{$\pm$} l | r @{$\pm$} l | r @{$\pm$} l}
\toprule
 
$Z\to e^+e^-$ & \multicolumn{2}{c|}{\Inc}   & \multicolumn{2}{c|}{\Hpt}   & \multicolumn{2}{c|}{\Col}  &\multicolumn{2}{c|}{\Btb} &\multicolumn{2}{c}{\Hht} \\
 
\midrule
\Zjets      & \numRF{1170846.5}{4}\phantom{0}  & \phantom{0}\numRF{49473.056}{2} & \numRF{6147.3934}{3}\phantom{0}  & \phantom{0}\numRF{310.93352}{2} & \numRF{2522.5307}{3}\phantom{0}  & \phantom{0}\numRF{121.26058}{2} & \numRF{2520.2496}{3}\phantom{0}  & \phantom{0}\numRF{153.47876}{2} & \numRF{18301.319}{3}\phantom{0}  & \phantom{0}\numRF{797.92362}{2}  \\
$t\bar{t}$   & \numRF{43397.211}{3}\phantom{0}  & \phantom{0}\numRF{1281.9149}{2} & \numRF{208.95027}{3}\phantom{0}  & \phantom{0}\numRF{16.329925}{2} & \numRF{135.82391}{3}\phantom{0}  & \phantom{0}\numRF{12.543188}{2} & \numRF{47.207162}{3}\phantom{0}  & \phantom{0}\numRF{7.4986662}{2} & \numRF{917.32899}{3}\phantom{0}  & \phantom{0}\numRF{41.337324}{2}  \\
Diboson         & \numRF{19533.437}{4}\phantom{0}  & \phantom{0}\numRF{752.48722}{2} & \numRF{428.18476}{3}\phantom{0}  & \phantom{0}\numRF{28.928236}{2} & \numRF{183.39648}{3}\phantom{0}  & \phantom{0}\numRF{15.713448}{2} & \numRF{166.74662}{3}\phantom{0}  & \phantom{0}\numRF{15.88231}{2} & \numRF{1007.5228}{4}\phantom{0}  & \phantom{0}\numRF{53.21198}{2}  \\
EW $Zjj$  & \numRF{13272.234}{4}\phantom{0}  & \phantom{0}\numRF{500.82358}{2} & \numRF{311.88656}{3}\phantom{0}  & \phantom{0}\numRF{23.081748}{2} & \numRF{102.27947}{3}\phantom{0}  & \phantom{0}\numRF{11.021392}{2} & \numRF{135.27504}{3}\phantom{0}  & \phantom{0}\numRF{13.711266}{2} & \numRF{789.11065}{3}\phantom{0}  & \phantom{0}\numRF{42.703701}{2}  \\
Single-top  & \numRF{2433.4204}{3}\phantom{0}  & \phantom{0}\numRF{159.28369}{2} & \numRF{27.937782}{3}\phantom{0}  & \phantom{0}\numRF{5.4540363}{2} & \numRF{14.033898}{3}\phantom{0}  & \phantom{0}\numRF{3.8028927}{2} & \numRF{9.7523204}{2}\phantom{0}  & \phantom{0}\numRF{3.1829567}{2} & \numRF{54.188443}{3}\phantom{0}  & \phantom{0}\numRF{8.1526252}{2}  \\
$Z \to \tau\tau$     & \numRF{514.71882}{3}\phantom{0}  & \phantom{0}\numRF{36.946337}{2} & \numRF{4.6185037}{2}\phantom{0}  & \phantom{0}\numRF{4.2259361}{2} & \numRF{1.5954175}{2}\phantom{0}  & \phantom{0}\numRF{2.1493511}{2} & \numRF{2.1629878}{2}\phantom{0}  & \phantom{0}\numRF{1.7401434}{2} & \numRF{10.608423}{3}\phantom{0}  & \phantom{0}\numRF{6.2032768}{2}  \\
\Wjets     & \numRF{92.721706}{2}\phantom{0}  & \phantom{0}\numRF{15.689759}{2} & \numRF{3.4323967}{2}\phantom{0}  & \phantom{0}\numRF{1.870008}{2} & \numRF{0.31661122}{1}\phantom{0}  & \phantom{0}\numRF{0.60557352}{1} & \numRF{2.9144832}{2}\phantom{0}  & \phantom{0}\numRF{1.7383324}{2} & \numRF{3.3764906}{2}\phantom{0}  & \phantom{0}\numRF{1.9079829}{2}  \\
\Vgamma   & \numRF{1412.8706}{4}\phantom{0}  & \phantom{0}\numRF{82.606767}{2} & \numRF{14.17179}{3}\phantom{0}  & \phantom{0}\numRF{4.2919927}{2} & \numRF{6.4904071}{2}\phantom{0}  & \phantom{0}\numRF{2.558192}{2} & \numRF{5.1001561}{2}\phantom{0}  & \phantom{0}\numRF{2.3259103}{2} & \numRF{34.145236}{3}\phantom{0}  & \phantom{0}\numRF{7.3131851}{2}  \\
\hline
Total predicted     & \numRF{1251503.1}{4}\phantom{0}  & \phantom{0}\numRF{51096.093}{2} & \numRF{7146.5755}{3}\phantom{0}  & \phantom{0}\numRF{346.51857}{2} & \numRF{2966.4669}{3}\phantom{0}  & \phantom{0}\numRF{134.71916}{2} & \numRF{2889.4083}{3}\phantom{0}  & \phantom{0}\numRF{170.60084}{2} & \numRF{21117.599}{3}\phantom{0}  & \phantom{0}\numRF{880.73121}{2}  \\
\hline
Data        & \multicolumn{2}{c|}{\num{1312145}} & \multicolumn{2}{c|}{\num{7539}} & \multicolumn{2}{c|}{\num{2955}} & \multicolumn{2}{c|}{\num{3231}} & \multicolumn{2}{c}{\num{21746}} \\
\bottomrule
\toprule
 
$Z\to\mumu$ & \multicolumn{2}{c|}{\Inc}   & \multicolumn{2}{c|}{\Hpt}   & \multicolumn{2}{c|}{\Col}  &\multicolumn{2}{c|}{\Btb} &\multicolumn{2}{c}{\Hht} \\
 
\midrule
\Zjets      & \numRF{1536589.6}{4}\phantom{0}  & \phantom{0}\numRF{62593.684}{2} & \numRF{6702.8684}{3}\phantom{0}  & \phantom{0}\numRF{298.98438}{2} & \numRF{2951.2812}{3}\phantom{0}  & \phantom{0}\numRF{131.50912}{2} & \numRF{2423.3685}{3}\phantom{0}  & \phantom{0}\numRF{124.67934}{2} & \numRF{23105.351}{4}\phantom{0}  & \phantom{0}\numRF{915.33966}{2}  \\
$t\bar{t}$   & \numRF{55400.213}{3}\phantom{0}  & \phantom{0}\numRF{1324.9909}{2} & \numRF{208.72172}{3}\phantom{0}  & \phantom{0}\numRF{15.744822}{2} & \numRF{141.64235}{3}\phantom{0}  & \phantom{0}\numRF{12.489069}{2} & \numRF{39.119321}{3}\phantom{0}  & \phantom{0}\numRF{6.5866596}{2} & \numRF{1057.7632}{4}\phantom{0}  & \phantom{0}\numRF{41.245023}{2}  \\
Diboson         & \numRF{24161.443}{4}\phantom{0}  & \phantom{0}\numRF{865.44537}{2} & \numRF{438.44338}{3}\phantom{0}  & \phantom{0}\numRF{27.369331}{2} & \numRF{197.66834}{3}\phantom{0}  & \phantom{0}\numRF{16.110116}{2} & \numRF{157.01301}{3}\phantom{0}  & \phantom{0}\numRF{14.276075}{2} & \numRF{1148.6807}{4}\phantom{0}  & \phantom{0}\numRF{54.700017}{2}  \\
EW $Zjj$  & \numRF{17015.421}{4}\phantom{0}  & \phantom{0}\numRF{578.61006}{2} & \numRF{327.78734}{3}\phantom{0}  & \phantom{0}\numRF{22.497184}{2} & \numRF{113.21448}{3}\phantom{0}  & \phantom{0}\numRF{11.505719}{2} & \numRF{134.43141}{3}\phantom{0}  & \phantom{0}\numRF{13.020826}{2} & \numRF{915.3224}{3}\phantom{0}  & \phantom{0}\numRF{45.256539}{2}  \\
Single-top  & \numRF{3109.8351}{3}\phantom{0}  & \phantom{0}\numRF{185.8685}{2} & \numRF{29.06222}{3}\phantom{0}  & \phantom{0}\numRF{5.5234989}{2} & \numRF{13.63139}{3}\phantom{0}  & \phantom{0}\numRF{3.7607661}{2} & \numRF{11.178037}{3}\phantom{0}  & \phantom{0}\numRF{3.4586459}{2} & \numRF{69.967067}{3}\phantom{0}  & \phantom{0}\numRF{9.1643058}{2}  \\
$Z \to \tau\tau$     & \numRF{460.42803}{3}\phantom{0}  & \phantom{0}\numRF{33.350229}{2} & \numRF{3.4916262}{2}\phantom{0}  & \phantom{0}\numRF{3.9923945}{2} & \numRF{1.1267798}{2}\phantom{0}  & \phantom{0}\numRF{2.3235953}{2} & \numRF{1.7920218}{2}\phantom{0}  & \phantom{0}\numRF{1.5098696}{2} & \numRF{8.7779676}{2}\phantom{0}  & \phantom{0}\numRF{5.4285186}{2}  \\
\Wjets     & \numRF{128.03794}{3}\phantom{0}  & \phantom{0}\numRF{13.581148}{2} & \numRF{1.8755905}{2}\phantom{0}  & \phantom{0}\numRF{1.3957072}{2} & \numRF{0.29847557}{1}\phantom{0}  & \phantom{0}\numRF{0.54771003}{1} & \numRF{1.5150899}{2}\phantom{0}  & \phantom{0}\numRF{1.2596683}{2} & \numRF{2.6737423}{2}\phantom{0}  & \phantom{0}\numRF{1.988691}{2}  \\
\Vgamma   & \numRF{1272.7255}{4}\phantom{0}  & \phantom{0}\numRF{90.224688}{2} & \numRF{2.483629}{2}\phantom{0}  & \phantom{0}\numRF{2.3755164}{2} & \numRP{0.000}{1}\phantom{0}  & \phantom{0}\numRF{0.7}{1} & \numRF{2.2210466}{2}\phantom{0}  & \phantom{0}\numRF{1.5022874}{2} & \numRF{22.352626}{3}\phantom{0}  & \phantom{0}\numRF{5.4936813}{2}  \\
\hline
Total predicted     & \numRF{1638137.7}{4}\phantom{0}  & \phantom{0}\numRF{64189.319}{2} & \numRF{7714.7336}{3}\phantom{0}  & \phantom{0}\numRF{330.14452}{2} & \numRF{3417.6005}{3}\phantom{0}  & \phantom{0}\numRF{144.81666}{2} & \numRF{2770.6386}{3}\phantom{0}  & \phantom{0}\numRF{137.14125}{2} & \numRF{26330.888}{3}\phantom{0}  & \phantom{0}\numRF{996.19239}{2}  \\
\hline
Data        & \multicolumn{2}{c|}{\num{1673057}} & \multicolumn{2}{c|}{\num{7896}} & \multicolumn{2}{c|}{\num{3372}} & \multicolumn{2}{c|}{\num{3059}} & \multicolumn{2}{c}{\num{26567}} \\
\bottomrule
 
\end{tabular}
}
\label{tab:backgrounds}
\end{table}


The \ttbar\ background is evaluated with a data-driven methodology. A \ttbar-enriched control region is constructed with the same event selection as the signal region, but with $e^\pm \mu^\mp$ final states instead of the same-flavour $\ee$ or $\mumu$ pairs. This control region contains only percent-level contributions from \Zjets, \Wjets, diboson, single-top and \Ztt\ events. The prediction for the \ttbar\ distributions in the signal region is obtained by multiplying the corresponding measured distributions in the control region (after subtracting the non-\ttbar\ contributions) by $ee/e\mu$ and $\mu\mu/e\mu$ scale factors \cite{STDM-2012-04}. These factors are computed bin by bin in the signal region for each distribution using simulation.
 
Diboson backgrounds are dominated by two contributions: semileptonic $WZ$ and $ZZ$ final states, and fully leptonic diboson final states where the decay products of a gauge boson are reconstructed as jets. The measured kinematics of the boson decay products, as well as the production of one or two additional jets, agrees with predictions from \SHERPA, as demonstrated in Refs.~\cite{STDM-2017-03, STDM-2017-24, STDM-2018-03, STDM-2016-15, STDM-2015-23} within the modelling uncertainties described in Section~\ref{sec:Systematics}.
The simulation of the EW \Zjj\ events done with \herwigNoSeven{}+\VBFNLO agrees with measurements performed in ($Z \to\ell^{+}\ell^{-}$)-enriched phase spaces~\cite{STDM-2017-27}. Due to their good performance in previous measurements, simulations are used to describe the diboson and EW \Zjj\ backgrounds in this analysis.
 
Multijet events are assessed with a data-driven approach using a template fit of the \mll\ distribution. The \mll\ template for this background is derived from data in a multijet-enriched control region, which is defined by either inverting or dropping the lepton selection requirements associated with isolation, identification and charge. The sub-percent contributions to the multijet template that do not originate from the multijet background are evaluated and subtracted using simulation. The fit is performed over the range $51~\GeV \leq \mll \leq 151~\GeV$. The contribution from multijet events in the analysis is then estimated in the invariant-mass interval of the signal region ($71~\GeV \leq \mll \leq 111~\GeV$).
The resulting fraction of multijet events is at the sub-percent level and so is neglected in this analysis.
 
Figure~\ref{fig:CP-Kinematics} shows the data and predicted event yields as a function of \ptlj\ in the electron channel and as a function of \mDRZj\ in the \hpt\ region in the muon channel. The \Shtte\ predictions agree with the data in general, but do not describe it precisely in the full range of the measurements. The distributions are discussed in more detail in Section~\ref{sec:Results}.
 
\begin{figure}[!ht]
\centering
\subfloat{
\includegraphics[width=0.50\textwidth]{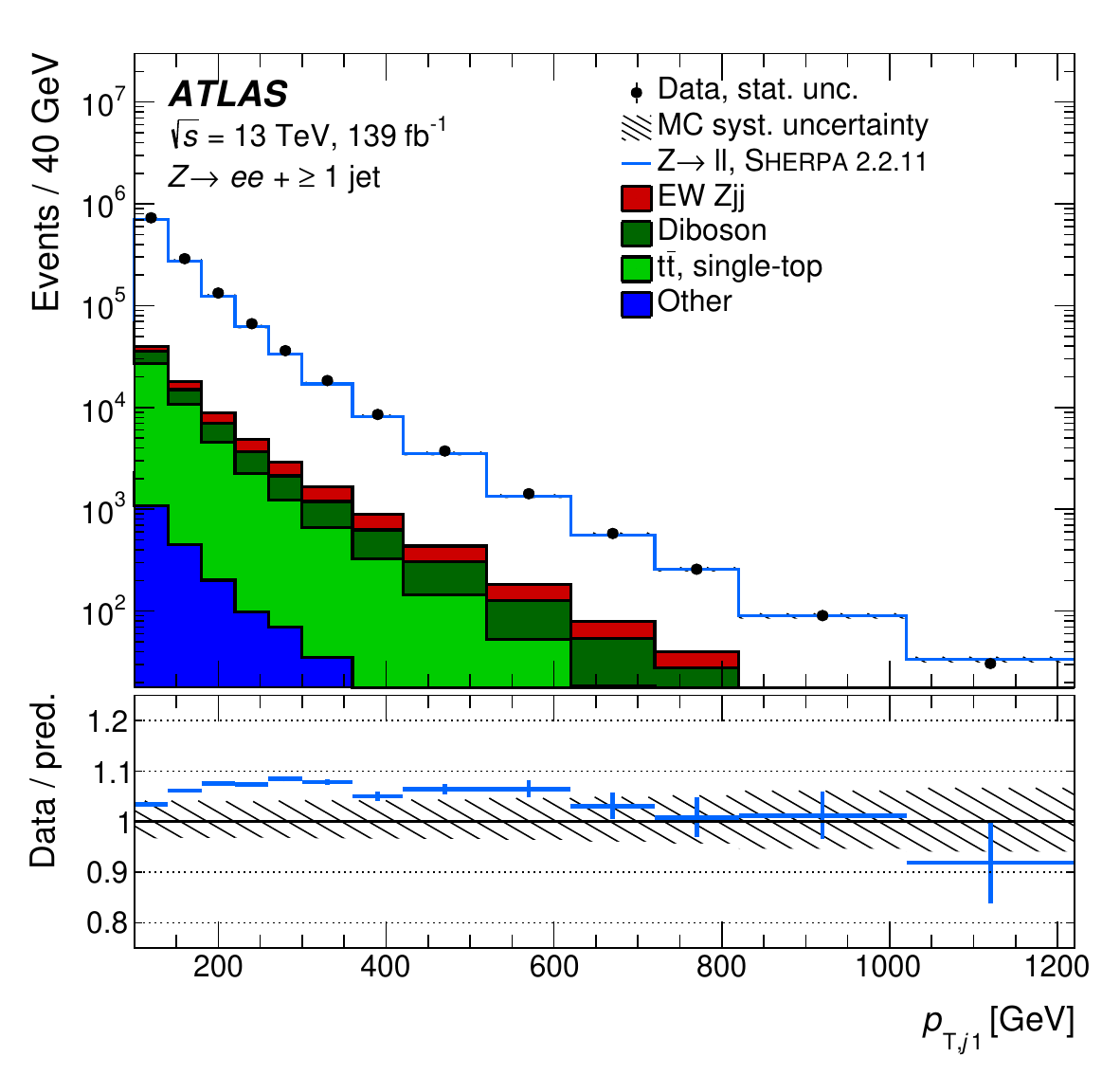}
}
\subfloat{
\includegraphics[width=0.50\textwidth]{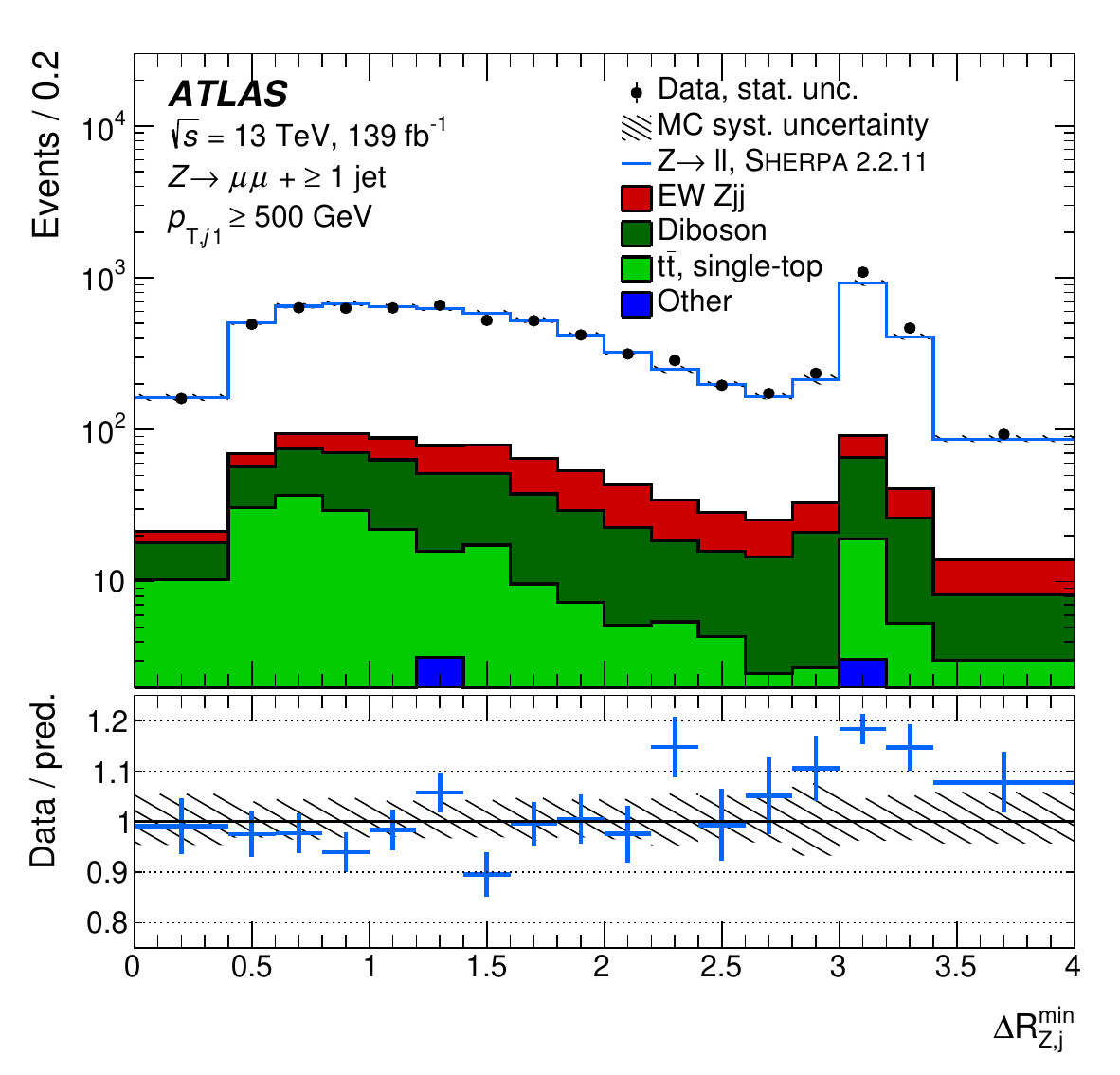}
}
\caption{Distributions of the leading-jet \pT\ in the \inc\ region in the electron channel (left) and the angular distance between the \ZBoson\ and the closest jet, $\mDRZj$, in the \hpt\ region in the muon channel (right). The signal and background samples are stacked to produce the figures. The \Wjets, $Z\to\tau^-\tau^+$ and \Vgamma\ processes are combined and labelled `Other'. The bottom panel shows the ratio of the data to the total prediction. Experimental uncertainties (described in Section~\ref{sec:Systematics}) for the signal and background distributions are combined in the hatched band, and the data statistical uncertainty is shown as error bars.}
\label{fig:CP-Kinematics}
\end{figure}


\section{Unfolding of detector effects}
\label{sec:Unfolding}

The cross-section measurements presented in this paper (see Section~\ref{sec:Introduction})  are performed within the fiducial acceptance region defined by the following requirements:
\begin{itemize}
\item Two same-flavour, opposite-charge leptons  with \pT $\geq 25~\GeV$ and $|\eta| < 2.5$
\item $71~\GeV \leq \mll \leq 111~\GeV$
\item At least one jet, where jets must have $\pT \geq 100~\GeV$ and $|y | < 2.5$.
\end{itemize}
The \hpt, \col, \btb, and \hht\ signal regions are defined in analogy to the detector-level definitions.

The cross sections are defined at particle level, corresponding to `dressed' electrons and muons. A dressed lepton is defined as the four-vector combination of a prompt lepton (that does not originate from the decay of a hadron or a $\tau$-lepton, or from a photon conversion) and all prompt photons within a surrounding cone of size $\Delta R =0.1$. The particle level also includes jets found by applying the anti-\kt\ algorithm with radius parameter $R = 0.4$ to final-state particles with decay length $c\tau > 10$~mm, excluding dressed \Zdash-boson decay products. Overlap removal is also applied at particle level: jets with $\pT \geq 30$~\GeV\ within $\Delta R = 0.2$ of a dressed lepton are removed, followed by the removal of leptons within $\Delta R = 0.4$ of the remaining jets. This overlap removal is applied at particle level in order to best match the detector response, especially in the \col\ region where the detector is not able to discriminate easily between nearby objects.

The fiducial cross sections are evaluated from the reconstructed kinematic observables for events that pass the selection described in Section~\ref{sec:Reconstruction}. The expected background components, as described in Section~\ref{sec:Backgrounds}, are subtracted from the distributions in data.
 
An iterative unfolding technique~\cite{DAgostini:1995} with two iterations, as implemented in the RooUnfold package~\cite{Adye:2011gm}, is used to unfold the background-subtracted data to the particle level, thereby accounting for the impact of detector inefficiencies and resolution~\cite{PERF-2018-01,PERF-2015-10,MUON-2018-03,PERF-2015-09,JETM-2018-05}. Before entering the iterative unfolding, the background-subtracted data are corrected for the expected fraction of events passing the detector-level selection but not the particle-level selection. The unfolding is carried out with the response matrices constructed from the \Shtte\ \Zjets\ samples. The unfolded event yields are divided by the integrated luminosity of the data sample to provide the final fiducial cross sections~\cite{STDM-2016-01}. The electron and muon channels are unfolded separately and then combined to measure the production cross section for a \Zboson boson decaying into a single charged-lepton flavour ($Z \to\ell^{+}\ell^{-}$).
 
The binning of all observables is optimised to keep the statistical uncertainty below $10\%$  and to maximize the  purity, or the fraction of reconstructed events where the reconstructed and truth values fall in the same bin.  The latter is kept above $60\%$, with typical values of  $70-90\%$. In order to mitigate the modelling uncertainty due to migration effects across the lower edge of the $\ptlj$ distribution, this observable is unfolded using two underflow bins within $60~\GeV \leq \ptlj \leq 100~\GeV$. In a similar fashion, an underflow bin is added for the \ptZ, \nht\ and jet multiplicity distributions for events where the leading jet does not pass the $\pt \geq 100$~\GeV\ selection but instead has $60~\GeV \leq \ptlj \leq 100~\GeV$. The unfolding of $\ptzj$ is performed in two dimensions using three bins for the \pT\ of the closest jet for each bin of $\ptzj$.


\section{Systematic uncertainties }
\label{sec:Systematics}
\subsection*{Theory modelling uncertainties}
Theoretical modelling uncertainties from the MC predictions are considered when unfolding the data and in the comparisons with the cross-section measurements.
 
Modelling uncertainties are taken into account by varying the QCD scales, the PDFs and, in the case of \Shtte, the virtual EW corrections. The effect of QCD scale uncertainties is defined by the envelope of variations resulting from changing the renormalisation (\muR) and factorisation (\muF) scales by factors of two with an additional constraint of $0.5\leq \muR / \muF \leq2$. Uncertainties due to the PDF parameterisation are evaluated using sets of PDF variations~\cite{Butterworth:2015oua}.
The PDF uncertainties also include a comparison with the nominal \MMHT[nnlo]~\cite{Harland-Lang:2014zoa-CUSTOM} PDF and the \CT[14nnlo]~\cite{Dulat:2015mca} PDFs. For \SHERPA\ \Zjets\ and diboson processes, the PDF uncertainties also include a consistent variation of \alphas\  in the PDF and in the hard scatter based on \NNPDFthreeNLO~\cite{Ball:2014uwa}. The prediction from \Shtte\ also considers uncertainties related to the NLO virtual EW corrections, derived from the envelope of the additive, multiplicative and exponentiated EW correction schemes~\cite{PMGR-2021-01}. The uncertainties associated with the virtual EW corrections are maximal and amount to $5\%$ where the EW corrections are largest: in the \btb\ region with large \ptZ\ and $\minDR \approx \pi$. In comparison, the effect of QCD scale uncertainties on \Shtte\ predictions ranges between 10$\%$ and 60$\%$, with average values near 25$\%$. The corresponding range in \MGfnlo\ is between 5$\%$ and 20$\%$. The differences between these two generators and their uncertainties are further explored in Ref.~\cite{PMGR-2021-01}. In  the \NNLOJET\ predictions, the QCD scale uncertainties are typically in the range between 5$\%$ and 10$\%$ and constitute the dominant systematic component. Due to computational limitations in the \NNLOJET\ program, the predictions do not include PDF uncertainties.

The diboson predictions used in the background subtraction from data include PDF and scale uncertainties. The EW \Zjj\ prediction includes the effects of scale, PDF and interference uncertainties, which amount to normalisation uncertainties of $9\%$, $2\%$ and 25\%--40\% respectively (see Section~\ref{sec:DataMCSamples}). The effects of scale and PDF uncertainties on the single-top predictions amount to a total normalisation uncertainty of about $4\%$, primarily from the normalisation to theory predictions at NNLO and NLO accuracies.
 
\subsection*{Systematic uncertainties in cross-section measurements}
Systematic uncertainties in the measured cross sections stem from experimental, MC-modelling and unfolding uncertainties. The uncertainties are propagated to the data cross sections by varying the subtracted background and the MC inputs to the unfolding procedure (response matrix, fraction of unmatched events, reconstruction efficiency). They are treated as being correlated over kinematic regions, over distributions of observables and, where applicable, over channels and between signal and background processes.
 
\textit{Experimental uncertainties specific to each leptonic final state ($Z\to e^+e^-$ and $Z\to\mumu$)}:  Systematic uncertainties in the lepton-candidate selection are related to the reconstruction, identification, isolation, and trigger~\cite{PERF-2018-01,PERF-2015-10}. Uncertainties in the lepton calibrations can cause changes in the acceptance, owing to the migration of events across the \pt\ threshold and the \mll\ boundaries. The uncertainties in the electron energy scale and resolution are taken into account~\cite{PERF-2017-01}, as are those related to the muon momentum scale, inner-detector and muon-spectrometer resolution, and sagitta-bias correction~\cite{PERF-2015-10}.
 
\textit{Experimental uncertainties common to the electron and muon final states}: Systematic uncertainties associated with jet reconstruction are addressed via jet-energy-scale (JES) variations in a 29-nuisance-parameter scheme and jet-energy-resolution (JER) variations in a 13-nuisance-parameter scheme~\cite{PERF-2016-04,PERF-2011-04}.
Imperfect modelling of the effects of pile-up leads to acceptance changes for different jet multiplicities. To assess this uncertainty, the average number of pile-up interactions is varied in simulation. The uncertainty in the combined 2015--2018 integrated luminosity is 1.7\% \cite{ATLAS-CONF-2019-021}, obtained using the LUCID-2 detector~\cite{LUCID2} for the primary luminosity measurements.
 
\textit{Modelling uncertainties}: Distribution-shape variations from PDF, scale and EW uncertainties in the \Shtte\ \Zjets\ simulation, computed as described above, are propagated to the unfolded cross sections via the response matrices and associated unmatched-events and efficiency corrections. Although the uncertainties in the simulation can be large, their effect on the cross-section measurement is minimised by the unfolding technique used. The systematic uncertainties in the modelling of background MC samples are propagated to the unfolded cross section via the background subtraction in the signal regions. The background-modelling uncertainty comes mainly from the diboson and EW \Zjj\ backgrounds.
 
\textit{Systematic uncertainties associated with the unfolding procedure}: Systematic uncertainties account for possible residual biases in the unfolding procedure, such as those due to the modelling of the signal events or the finite bin width used in each distribution. The limited size of a simulation sample can also create biases in the distribution used in the unfolding procedure. The following uncertainties from the unfolding procedure are considered:
\begin{itemize}
\item The statistical uncertainties of the MC inputs to the unfolding procedure are propagated to the unfolded cross sections with a `toy' simulation method based on  1000 ensembles (pseudo-experiments) of unfolding inputs.
 
\item The effects of the mismodelling of the data by the MC simulation on the results of unfolding procedure are derived by reweighting the \Shtte\ \Zjets\ MC simulation at particle level for each unfolded observable, such that the MC simulation distribution matches the background-subtracted data at the reconstruction level. The reweighted MC simulation is unfolded with the non-reweighted response matrix and the uncertainty is obtained by comparing the unfolded result against the reweighted distribution at particle level (non-closure).
 
\item An additional uncertainty is derived to account for more subtle differences between the \Shtte\ and \MGflo\ generators (e.g.\ hadronisation models, additional soft objects, distributions in other kinematic dimensions) which are not accounted for by the previous method. A non-closure test is performed where the \MGflo\ samples are first reweighed to match \Shtte\ particle-level distributions for each observable in turn and subsequently unfolded with \Shtte. The uncertainty is evaluated by comparing the unfolded result and the reweighted distribution at particle level.
\end{itemize}

The total fractional uncertainties of the unfolded differential cross sections in \ptlj\ and \mDRZj\ for the combined $Z \to\ell^{+}\ell^{-}$  measurement, performed as detailed in Section~\ref{sec:Results}, are shown in Figure~\ref{fig:UncPlot}. Table~\ref{tab:total-unc} shows the breakdown of the total relative statistical and systematic uncertainties in the measured integrated cross sections for \Zjets\ production in the five kinematic search regions. In the \inc\ and \hht\ regions, jet uncertainties dominate with a relative contribution of $2.6\%$ and $2.8\%$, respectively. In the \hpt\ region and its \col\ and \btb\ subregions the jet and data statistical uncertainties are largest, with relative contributions of $3.2\%$ and $2.1\%$ in the \hpt\ region, respectively,  $2.8\%$ and $2.9\%$ in the \col\ region, and $3.6\%$ and $2.7\%$ in the \btb\ region.

\begin{figure}[!ht]
\centering
\subfloat{
\includegraphics[width=0.50\textwidth]{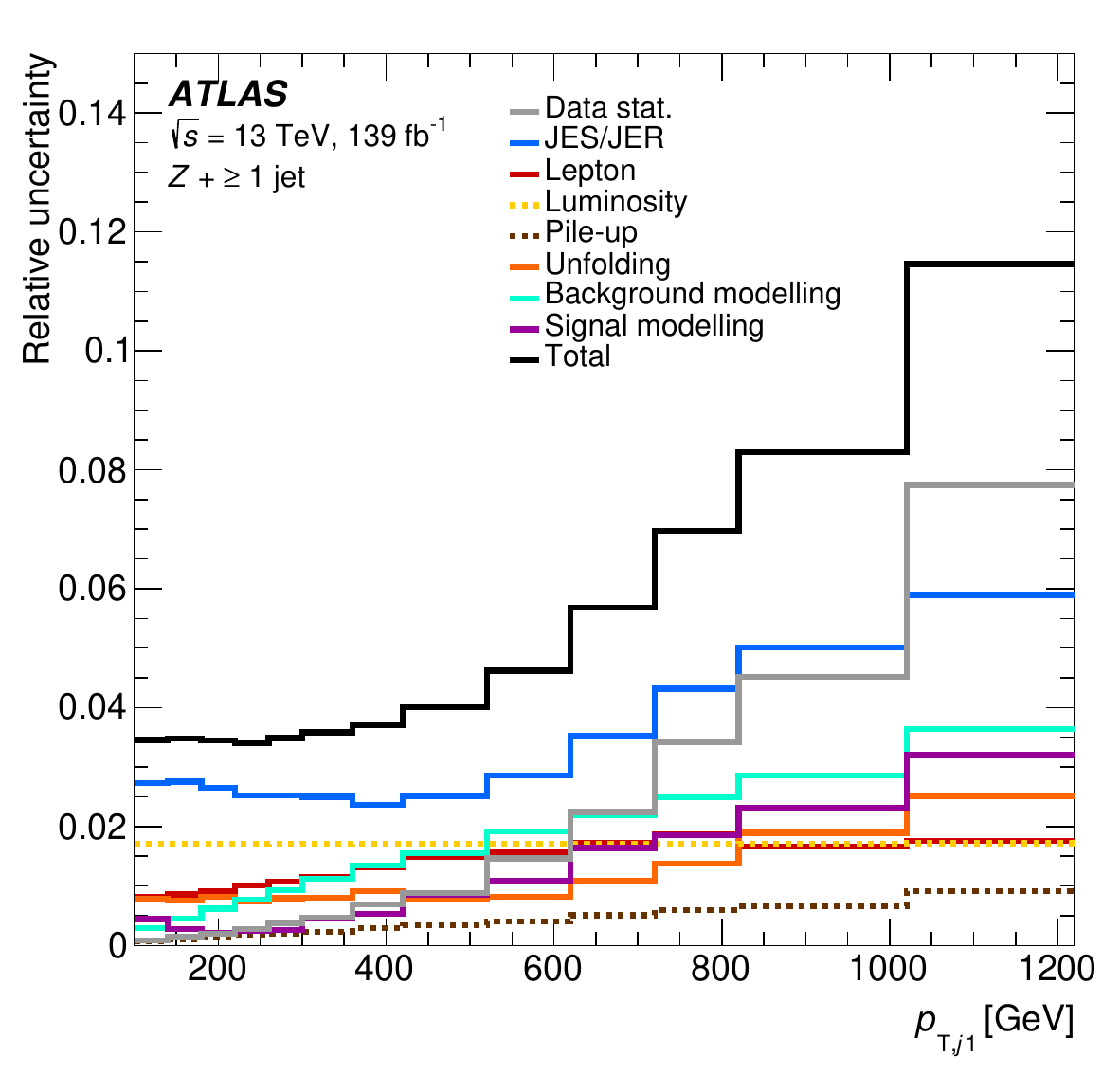}
}
\subfloat{
\includegraphics[width=0.50\textwidth]{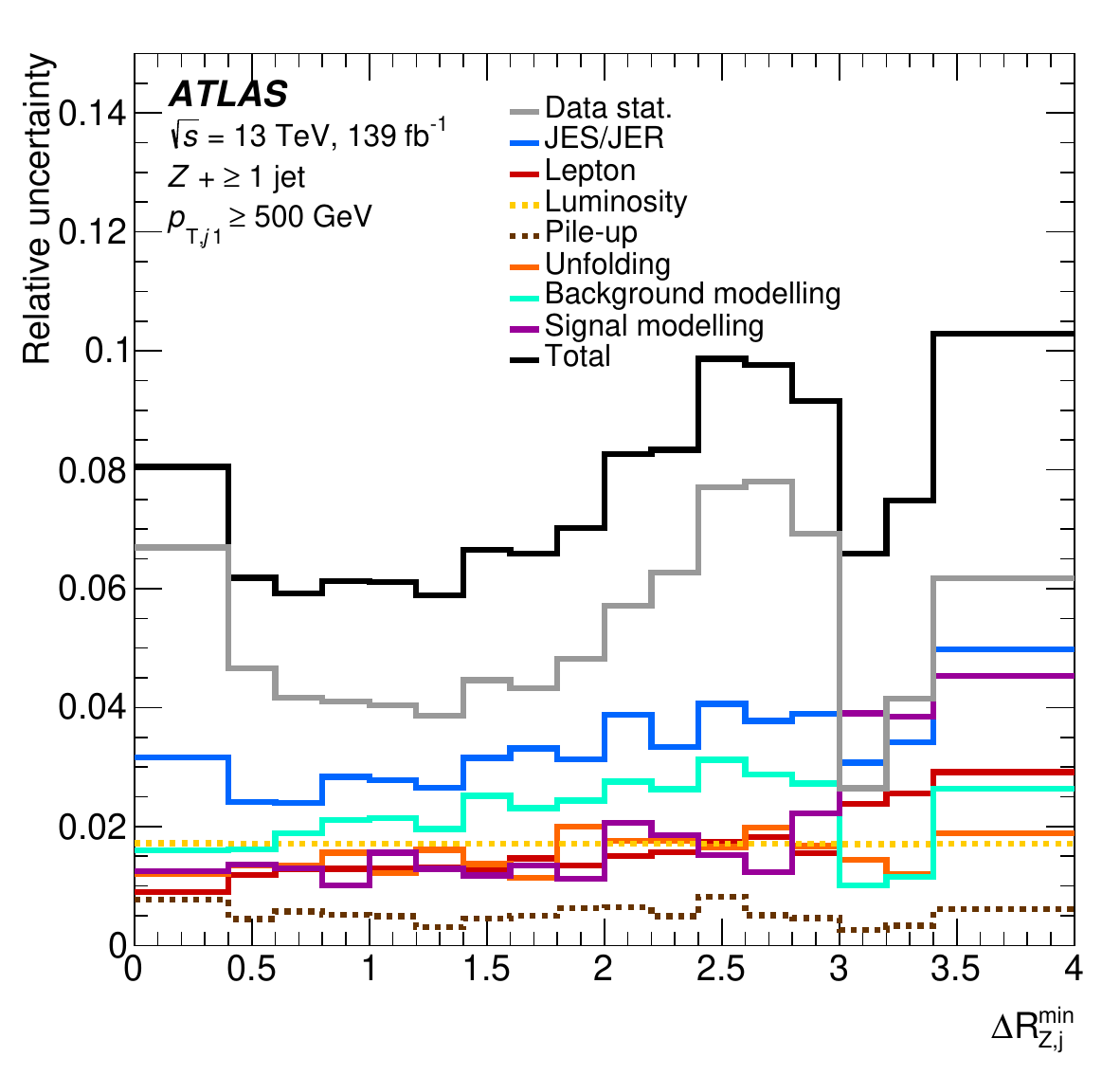}
}
\caption{Fractional uncertainties in the measured differential cross-sections in \ptlj\ in the \inc\ region (left) and in \mDRZj\ in the \hpt\ region (right) from the combined $Z \to\ell^{+}\ell^{-}$  measurement.}
\label{fig:UncPlot}
\end{figure}

\def\arraystretch{1.1}
\begin{table}[h]
\begin{center}
\caption{
Relative statistical and systematic uncertainties (in \%) in the measured integrated cross sections for \Zjets\ production in the five search regions, computed by integrating over the respective jet-multiplicity distribution.
\label{tab:total-unc}
}
\begin{footnotesize}
\begin{tabular}{ l | c | c | c | c | c }
\toprule
Uncertainty source [\%] & \Inc\   & \Hpt\  & \Col\  & \Btb\ & \Hht\  \\
\midrule
JES/JER              & \numRP{2.63366}{1} & \numRP{3.18128}{1} & \numRP{2.80625}{1} & \numRP{3.61022}{1} & \numRP{2.77058}{1} \\
Lepton                     & \numRP{0.870807}{1} & \numRP{1.61691}{1} & \numRP{1.37177}{1} & \numRP{2.01945}{1} & \numRP{1.1482}{1} \\
Luminosity               & \numRP{1.70166}{1} & \numRP{1.70697}{1} & \numRP{1.70933}{1} & \numRP{1.70564}{1} & \numRP{1.7067}{1} \\
Pile-up                  & \numRP{0.1111}{1} & \numRP{0.443335}{1} & \numRP{0.449973}{1} & \numRP{0.406108}{1} & \numRP{0.354038}{1} \\
Unfolding                & \numRP{0.506477}{1} & \numRP{1.02589}{1} & \numRP{1.08927}{1} & \numRP{1.40203}{1} & \numRP{0.81972}{1} \\
Background modelling     & \numRP{0.455076}{1} & \numRP{2.03349}{1} & \numRP{2.0334}{1} & \numRP{1.92054}{1} & \numRP{1.74976}{1} \\
Signal modelling         & \numRP{0.489737}{1} & \numRP{1.16052}{1} & \numRP{1.09705}{1} & \numRP{1.08907}{1} & \numRP{1.10456}{1} \\
\hline
Total syst.\ uncertainty  & \numRP{3.38148}{1} & \numRP{4.81593}{1} & \numRP{4.42059}{1} & \numRP{5.25365}{1} & \numRP{4.22249}{1} \\
Data stat.\ uncertainty   & \numRP{0.106554}{1} & \numRP{2.07792}{1} & \numRP{2.90132}{1} & \numRP{2.67605}{1} & \numRP{1.15283}{1} \\
\hline
Total uncertainty        & \numRP{3.38447}{1} & \numRP{5.26318}{1} & \numRP{5.31004}{1} & \numRP{5.94892}{1} & \numRP{4.39428}{1} \\
\bottomrule
 
\end{tabular}
\end{footnotesize}
\end{center}
\end{table}


\section{Results}
\label{sec:Results}
 
The integrated and differential fiducial cross sections are measured in the electron and muon channels separately, and the compatibility of the results from the two channels is tested. These results are then combined using the Best Linear Unbiased Estimate (BLUE) method~\cite{Lyons:1988rp}. In the combination, all systematic uncertainties except the lepton-related experimental uncertainties are treated as correlated between the electron and muon channels. Data and MC statistical uncertainties are treated as uncorrelated between channels. The combined measurements are consistent with both individual decay channels for the full set of observables. In general, the uncertainties in the measured cross sections are dominated by the systematic uncertainties and are smaller than the uncertainties in the predictions, except for \njnnlo, which matches or exceeds the precision of the measurements in some kinematic regions.
 
Integrated fiducial cross sections for \Zjets\ production are evaluated in the \inc, \hpt, \col, \btb\ and \hht\ signal regions (see Section~\ref{sec:Unfolding}) by summing over the respective unfolded jet-multiplicity distributions. The measured cross sections are compared with the predictions from \Shtte, \MGfnlo, \Shtto\ and with NNLO and NLO predictions from \NNLOJET\ in Table~\ref{total-xs} and in Figure~\ref{fig:xs_integrated}. The prediction from \Shtte\ uniquely includes NLO virtual EW corrections (see Section~\ref{sec:DataMCSamples}). When these virtual corrections are removed from the \Shtte\ prediction, its total cross sections for the \inc, \hpt, \col, \btb\ and \hht\ regions increase by $0.065\%$, $6.9\%$, $3.8\%$, $11\%$ and $3.0\%$, respectively. The cross sections predicted by the three generators and the \NNLOJET\ predictions agree with the measured values within the theory uncertainties.

\def\arraystretch{1.5}
\begin{table}[!htbp]
\caption{
Measured integrated fiducial cross sections for \Zjets\ production in the five signal regions and predictions from \Shtte, \MGfnlo, \Shtto, \njnnlo\ and \njnlo. Systematic uncertainties in the measured and predicted cross sections are calculated as described in Section~\ref{sec:Systematics}.
\label{total-xs}
}
\scriptsize
\begin{center}
\begin{tabular}{l| r l l l r}
\toprule
\multicolumn{5}{c}{\Inc\ $\Zjets$}\\
\hline
Data & \numRP{13.9045}{2} \phantom{0}&\phantom{0} $\pm$ \numRP{0.0148158}{2} (stat) \phantom{0}&\phantom{0} $\pm$ \numRP{0.470178}{2} (syst) & & pb \\
\hline
\Shtte & \numRP{13.2894}{1}\phantom{0} \phantom{0}&\phantom{0} $_{-\numRP{0.165659}{1}}^{+\numRP{0.243326}{1}}$ (PDF) \phantom{0}&\phantom{0} $_{-\numRP{1.80141}{1}}^{+\numRP{3.05905}{1}\phantom{0}}$ (Scale) \phantom{0}&\phantom{0} $ \pm \leq 0.1 $ (EW) & pb \\
\MGfnlo & \numRP{14.4531}{1}\phantom{0} \phantom{0}&\phantom{0} $_{-\numRP{0.0596007}{1}}^{+\numRP{0.136277}{1}}$ (PDF) \phantom{0}&\phantom{0} $_{-\numRP{1.23913}{1}}^{+\numRP{0.841971}{1}\phantom{0}}$ (Scale) & & pb \\
\Shtto & \numRP{13.7596}{1}\phantom{0} \phantom{0}&\phantom{0} $_{-\numRP{0.473338}{1}}^{+\numRP{0.507998}{1}}$ (PDF) \phantom{0}&\phantom{0} $_{-\numRP{3.42663}{1}}^{+\numRP{5.20082}{1}\phantom{0}}$ (Scale) & & pb \\
\njnnlo & \numRP{13.8282}{2} \phantom{0}& &\phantom{0} $_{-\numRP{0.27039}{2}}^{+\numRP{0.183804}{2}}$ (Scale) & & pb \\
\njnlo & \numRP{13.4507}{1} \phantom{0}\phantom{0}& &\phantom{0} $_{-\numRP{0.940599}{1}}^{+\numRP{1.09396}{1}\phantom{0}}$ (Scale) & & pb \\
\midrule
\multicolumn{5}{c}{\Hpt: $\ptlj \geq 500$~\GeV}\\
\hline
Data & \numRP{72.3261}{1} \phantom{0}&\phantom{0} $\pm$ \numRP{1.50288}{1}\phantom{0} (stat) \phantom{0}&\phantom{0} $\pm$ \numRP{3.48317}{1}\phantom{0} (syst) & & fb \\
\hline
\Shtte & \numRP{69.0908}{0}\phantom{.}\phantom{0} \phantom{0}&\phantom{0} $_{-\numRP{1.37719}{0}}^{+\numRP{1.80732}{0}\phantom{.}\phantom{0}}$ (PDF) \phantom{0}&\phantom{0} $_{-\numRP{16.7888}{0}}^{+\numRP{28.4267}{0}\phantom{.}}$ (Scale) \phantom{0}&\phantom{0} $_{-\numRP{2.28015}{0}}^{+\numRP{2.28015}{0}\phantom{.}}$ (EW) & fb \\
\MGfnlo & \numRP{78.287}{0}\phantom{.}\phantom{0} \phantom{0}&\phantom{0} $_{-\numRP{0.664618}{0}}^{+\numRP{3.62831}{0}\phantom{.}\phantom{0}}$ (PDF) \phantom{0}&\phantom{0} $_{-\numRP{12.0617}{0}\phantom{.}}^{+\numRP{9.16261}{0}\phantom{.}}$ (Scale) & & fb \\
\Shtto & \numRP{94.8483}{0}\phantom{.}\phantom{0} \phantom{0}&\phantom{0} $_{-\numRP{3.26293}{0}}^{+\numRP{3.95246}{0}\phantom{.}\phantom{0}}$ (PDF) \phantom{0}&\phantom{0} $_{-\numRP{26.3435}{0}}^{+\numRP{40.435}{0}\phantom{.}}$ (Scale) & & fb \\
\njnnlo & \numRP{75.5882}{0}\phantom{.}\phantom{0} \phantom{0}& &\phantom{0} $_{-\numRP{12.49}{0}}^{+\numRP{9.54897}{0}\phantom{.}}$ (Scale) & & fb \\
\njnlo & \numRP{71.0916}{0}\phantom{.}\phantom{0} \phantom{0}& &\phantom{0} $_{-\numRP{10.722}{0}}^{+\numRP{13.7479}{0}\phantom{.}}$ (Scale) & & fb \\
\midrule
\multicolumn{5}{c}{\Col: \Hpt\ and $\minDR \leq 1.4$}\\
\hline
Data & \numRP{27.906}{1} \phantom{0}&\phantom{0} $\pm$ \numRP{0.809642}{1}\phantom{0} (stat) \phantom{0}&\phantom{0} $\pm$ \numRP{1.23361}{1}\phantom{0} (syst) & & fb \\
\hline
\Shtte & \numRP{28.3206}{0}\phantom{.}\phantom{0} \phantom{0}&\phantom{0} $_{-\numRP{0.717432}{0}}^{+\numRP{0.874118}{0}\phantom{.}\phantom{0}}$ (PDF) \phantom{0}&\phantom{0} $_{-\numRP{8.26322}{0}}^{+\numRP{14.3331}{0}\phantom{.}}$ (Scale) \phantom{0}&\phantom{0} $ \pm \leq 1 $ (EW) & fb \\
\MGfnlo & \numRP{29.5937}{1} \phantom{0}&\phantom{0} $_{-\numRP{0.2521}{1}}^{+\numRP{1.27708}{1}}$ (PDF) \phantom{0}&\phantom{0} $_{-\numRP{4.27901}{1}}^{+\numRP{3.12548}{1}}$ (Scale) & & fb \\
\Shtto & \numRP{39.4541}{0}\phantom{.}\phantom{0} \phantom{0}&\phantom{0} $_{-\numRP{1.35447}{0}}^{+\numRP{1.61402}{0}\phantom{.}\phantom{0}}$ (PDF) \phantom{0}&\phantom{0} $_{-\numRP{11.3472}{0}}^{+\numRP{18.2152}{0}\phantom{.}}$ (Scale) & & fb \\
\njnnlo & \numRP{27.0113}{1} \phantom{0}& &\phantom{0} $_{-\numRP{7.23885}{1}}^{+\numRP{5.74364}{1}}$ (Scale) & & fb \\
\njnlo & \numRP{24.099}{1} \phantom{0}& &\phantom{0} $_{-\numRP{5.09163}{1}}^{+\numRP{7.02539}{1}}$ (Scale) & & fb \\
\midrule
\multicolumn{5}{c}{\Btb: \Hpt\ and $\minDR \geq 2.0$}\\
\hline
Data & \numRP{31.6117}{1} \phantom{0}&\phantom{0} $\pm$ \numRP{0.845944}{1}\phantom{0} (stat) \phantom{0}&\phantom{0} $\pm$ \numRP{1.66077}{1}\phantom{0} (syst) & & fb \\
\hline
\Shtte & \numRP{28.0769}{1} \phantom{0}&\phantom{0} $_{-\numRP{0.340563}{1}}^{+\numRP{0.552391}{1}}$ (PDF) \phantom{0}&\phantom{0} $_{-\numRP{4.89878}{1}}^{+\numRP{7.87002}{1}}$ (Scale) \phantom{0}&\phantom{0} $_{-\numRP{1.39058}{1}}^{+\numRP{1.39058}{1}}$ (EW) & fb \\
\MGfnlo & \numRP{34.3756}{1} \phantom{0}&\phantom{0} $_{-\numRP{0.335925}{1}}^{+\numRP{1.6438}{1}}$ (PDF) \phantom{0}&\phantom{0} $_{-\numRP{5.56958}{1}}^{+\numRP{4.56203}{1}}$ (Scale) & & fb \\
\Shtto & \numRP{38.4972}{0}\phantom{.}\phantom{0} \phantom{0}&\phantom{0} $_{-\numRP{1.33496}{0}}^{+\numRP{1.63826}{0}\phantom{.}\phantom{0}}$ (PDF) \phantom{0}&\phantom{0} $_{-\numRP{10.2238}{0}\phantom{.}}^{+\numRP{14.7207}{0}\phantom{.}}$ (Scale) & & fb \\
\njnnlo & \numRP{35.2816}{1} \phantom{0}& &\phantom{0} $_{-\numRP{2.43932}{1}}^{+\numRP{1.93832}{1}}$ (Scale) & & fb \\
\njnlo & \numRP{35.9928}{1} \phantom{0}& &\phantom{0} $_{-\numRP{3.30793}{1}}^{+\numRP{3.49491}{1}}$ (Scale) & & fb \\
\midrule
\multicolumn{5}{c}{\Hht: $\htj \geq 600$~\GeV}\\
\hline
Data & \numRP{225.987}{1} \phantom{0}&\phantom{0} $\pm$ \numRP{2.60524}{1}\phantom{0} (stat) \phantom{0}&\phantom{0} $\pm$ \numRP{9.54226}{1}\phantom{0} (syst) & & fb \\
\hline
\Shtte & \numRF{221.053}{2}\phantom{.}\phantom{0} \phantom{0}&\phantom{0} $_{-\numRF{10.4038}{1}}^{+\numRF{11.6619}{1}\phantom{.}}$ (PDF) \phantom{0}&\phantom{0} $_{-\numRF{61.9443}{1}}^{+\numRF{109.767}{2}}$ (Scale) \phantom{0}&\phantom{0} $ \pm \leq 10 $ (EW) & fb \\
\MGfnlo & \numRP{246.536}{0}\phantom{.}\phantom{0} \phantom{0}&\phantom{0} $_{-\numRP{1.59146}{0}\phantom{.}\phantom{0}}^{+\numRP{9.80537}{0}}$ (PDF) \phantom{0}&\phantom{0} $_{-\numRP{36.951}{0}}^{+\numRP{30.2417}{0}\phantom{0}}$ (Scale) & & fb \\
\Shtto & \numRF{282.981}{2}\phantom{.}\phantom{0} \phantom{0}&\phantom{0} $_{-\numRF{10.4226}{1}}^{+\numRF{11.286}{1}\phantom{.}}$ (PDF) \phantom{0}&\phantom{0} $_{-\numRF{79.9319}{1}}^{+\numRF{125.689}{2}}$ (Scale) & & fb \\
\njnnlo & \numRP{223.43}{0}\phantom{.}\phantom{0} \phantom{0}& &\phantom{0} $_{-\numRP{47.1156}{0}}^{+\numRP{42.5214}{0}\phantom{0}}$ (Scale) & & fb \\
\njnlo & \numRP{167.597}{0}\phantom{.}\phantom{0} \phantom{0}& &\phantom{0} $_{-\numRP{32.9246}{0}}^{+\numRP{44.639}{0}\phantom{0}}$ (Scale) & & fb \\
\bottomrule
\end{tabular}
\end{center}
\end{table}


\begin{figure*}[ht!]
\centering
\includegraphics[width=1.\textwidth]{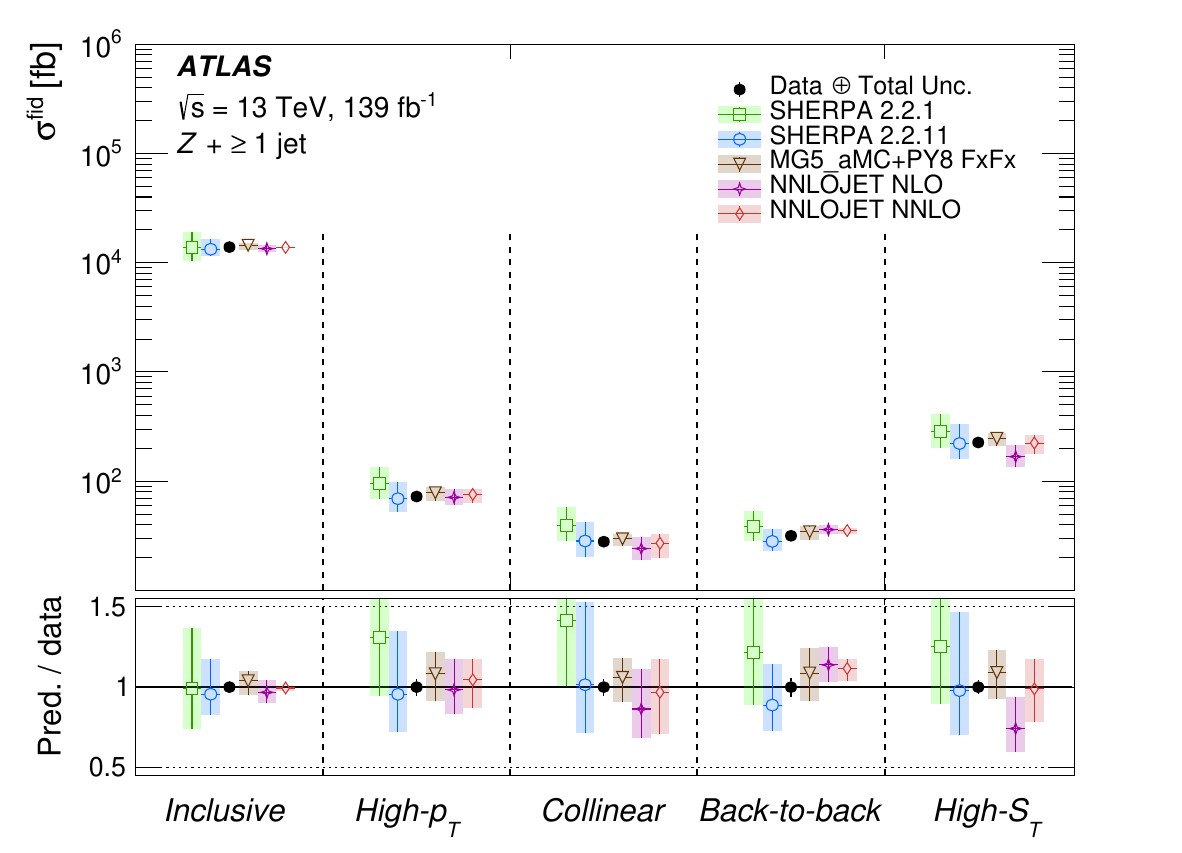}
\caption{
Summary of integrated fiducial cross-section results. The measured cross sections are shown with black points and the error bars represent the total uncertainty. Data are compared with predictions from MC generators and fixed-order calculations. The uncertainties in the predictions are found by adding in quadrature the uncertainties from the variations of the PDF (excluding \NNLOJET\ predictions), QCD scales and, for \Shtte, virtual EW contributions, as explained in Section~\ref{sec:Systematics}.
}
\label{fig:xs_integrated}
\end{figure*}

Differential cross sections are measured and compared in Figures~\ref{fig:xs_withNLO_1}--\ref{fig:xs_withNLO_7} with predictions from \Shtto, \MGflo, and the next-generation MC generators \Shtte\ and \MGfnlo, and with NLO and NNLO \Zjets\ calculations from \NNLOJET. In general, \njnnlo\ and \MGfnlo\ provide the most precise predictions.

The \Zdash-boson and jet transverse momenta (two correlated quantities) are fundamental observables of the \Zjets\ process and probe pertubative QCD over a wide range of scales. Moreover, understanding the kinematics of jets in events with vector bosons produced in association with several jets is essential for the modelling of backgrounds for other SM processes and searches beyond the SM. Figure~\ref{fig:xs_withNLO_1} shows the differential cross section as a function of \ptZ\ and \ptlj. The high-\ptZ\ region is dominated by the back-to-back topology and receives significant negative corrections due to EW effects.
In contrast, events with a \hpt\ jet typically result in both back-to-back and collinear topologies.  The \Shtto\ and \MGflo\ generators predict a harder \ptlj\ distribution than seen in the data, resulting in an overestimation of the cross section for high \ptlj. In contrast, \Shtte\  and \MGfnlo\ show significantly better modelling of the \ptlj\ spectrum and are also in good agreement with the measured \ptZ\ spectrum. The smaller cross section from \Shtte\ relative to \Shtto\ in the \hpt\ region is attributed to the improved matching scheme with a different treatment of unordered histories~\cite{PMGR-2021-01}. The prediction from \MGfnlo\ models the data more precisely, due to the inclusion of NLO matrix elements with three partons.
The \njnnlo\ predictions describe the data very precisely, except for very large values of \ptZ\ (and \ptlj), where negative NLO virtual EW corrections of 10\%--20\% are expected and the QCD-only calculation overestimates the data.

\begin{figure*}[t]
\centering
\subfloat{
\includegraphics[width=.50\textwidth]{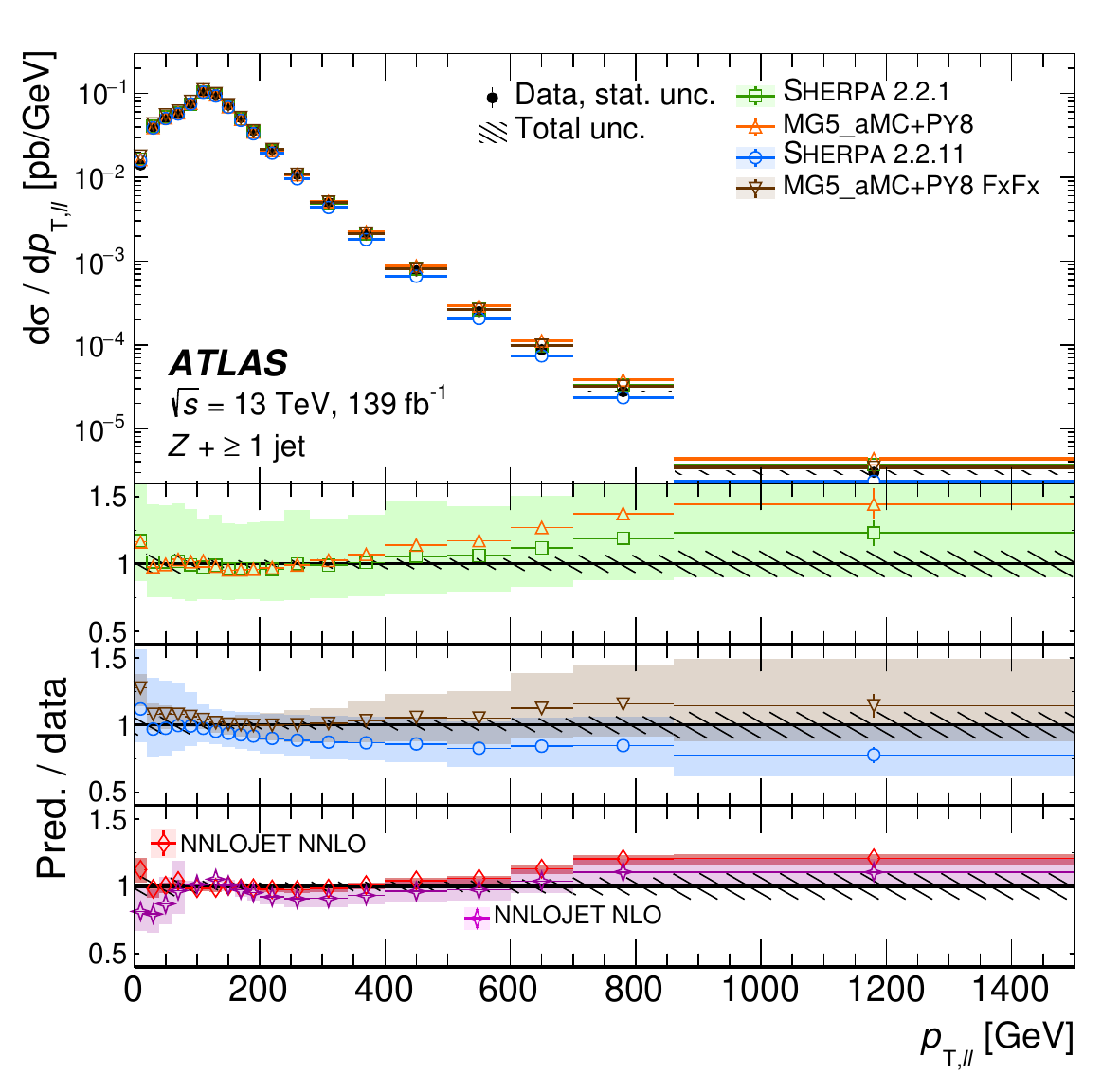}
}
\subfloat{
\includegraphics[width=.50\textwidth]{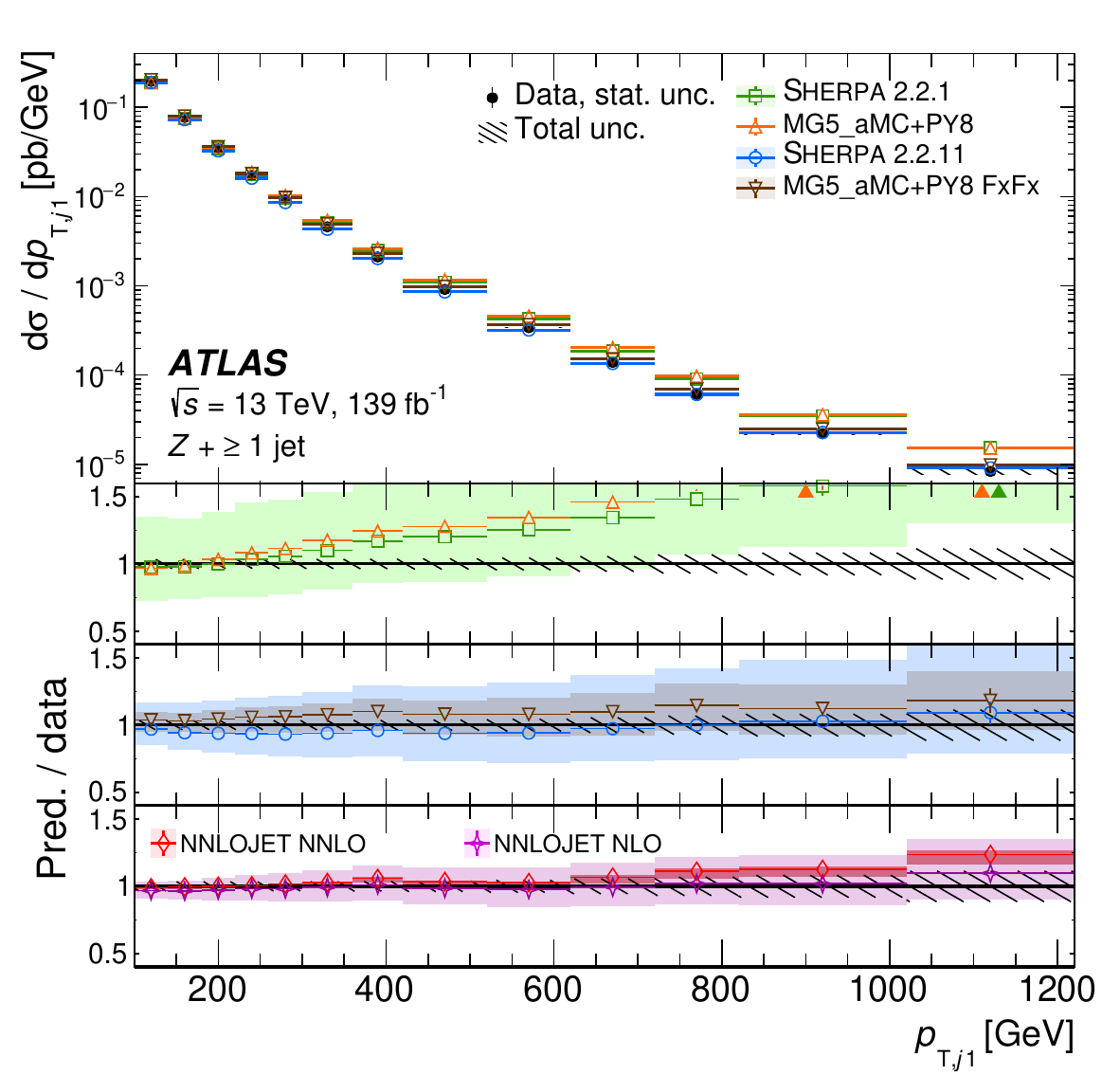}
}
\caption{
Differential cross section as a function of the transverse momentum of the \ZBoson\ (left) and of the transverse momentum of the leading jet (right) in the \inc\ region. The unfolded data are shown with the black points where the statistical uncertainty is given as an error bar and the total uncertainty as a hatched region. The data are compared with predictions from MC generators and fixed-order calculations. In the ratio panels, the error bars correspond to the statistical uncertainty of the prediction and solid triangles indicate that the prediction is outside the vertical-axis range, while the total uncertainty of the unfolded data is represented as the hatched region. The uncertainties in the predictions, dominated by the scale uncertainties, are shown only in the ratio panels, except for \MGflo\ which is not included. They are found by adding in quadrature the uncertainties from the variations of the PDF (excluding \NNLOJET\ predictions), QCD scales and, for \Shtte, virtual EW contributions, as explained in Section~\ref{sec:Systematics}.
}
\label{fig:xs_withNLO_1}
\end{figure*}

Jet-multiplicity distributions provide an excellent probe of QCD. Whereas the \Zonejet\ bin is most sensitive to PDF effects, those with higher jet multiplicities are more sensitive to perturbative QCD effects~\cite{scaling}.
Jet-multiplicity distributions also probe the validity of predictions in the presence of jet vetoes, which are frequently used in searches that require a specific number of jets in the selection. These vetoes create additional logarithmic terms, which are not explicitly included in the theoretical predictions presented in this paper.
Figure~\ref{fig:xs_withNLO_2} shows the differential cross section as a function of the jet multiplicity in the \inc\ and \hpt\ regions. As expected~\cite{scaling}, the jet multiplicity in the \inc\ phase space follows a downward staircase pattern, whereas in the \hpt\ phase space, the cross section increases between 1-jet and 2-jet events followed by a downward pattern for higher jet multiplicities.  While \Shtto\ and \MGflo\ tend to overestimate the cross section for higher jet multiplicities, \Shtte\  and \MGfnlo\ agree with the data for both the \inc\ and \hpt\ regions, the latter generator at a higher level of precision. The \NNLOJET\ predictions agree with the data in the \inc\ and \hpt\ phase spaces at high precision when it is expected from the order of the calculation, and at lower precision when the order of the calculation is exceeded.
 
\begin{figure*}[t]
\centering
\subfloat{
\includegraphics[width=.50\textwidth]{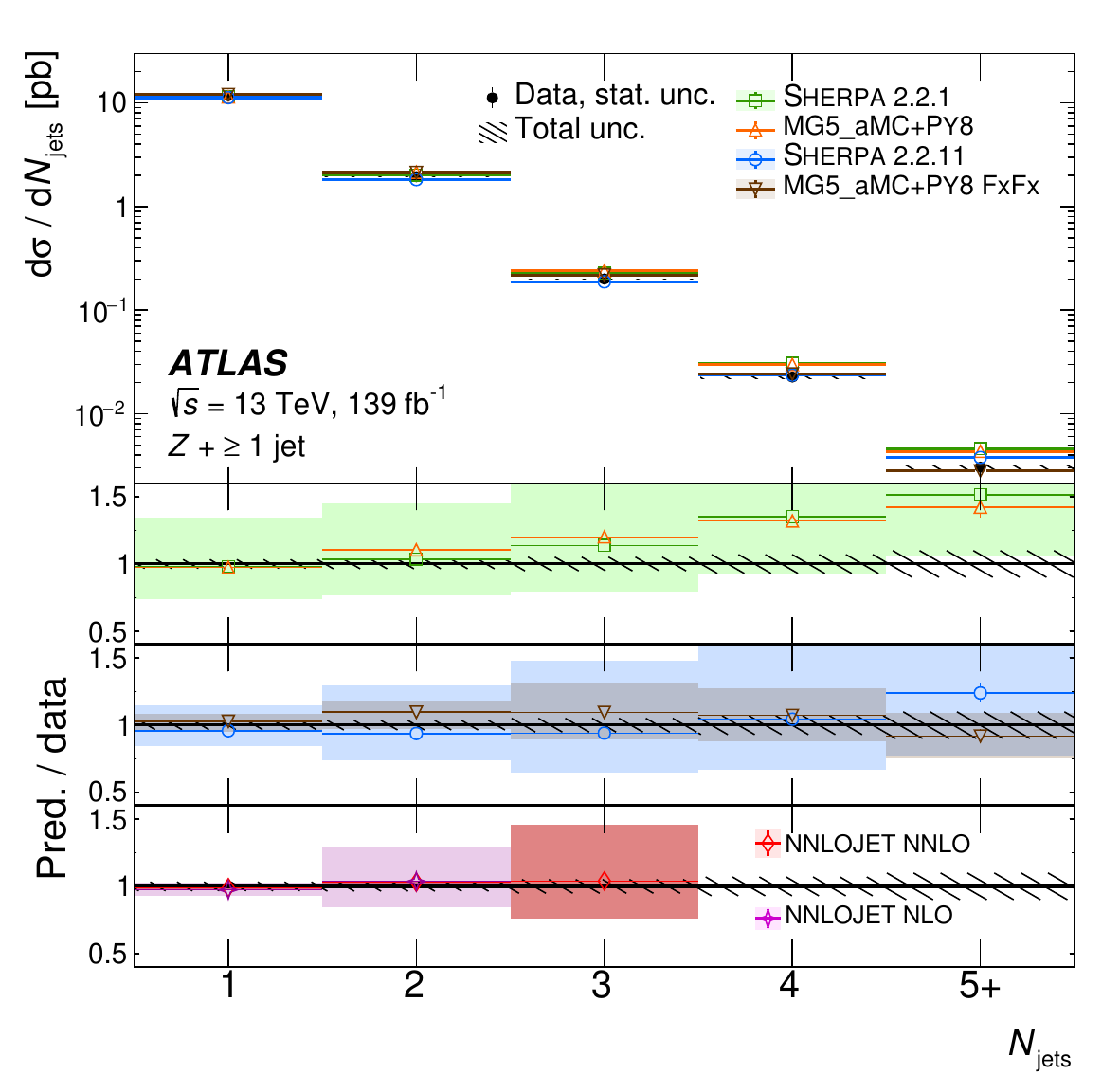}
}
\subfloat{
\includegraphics[width=.50\textwidth]{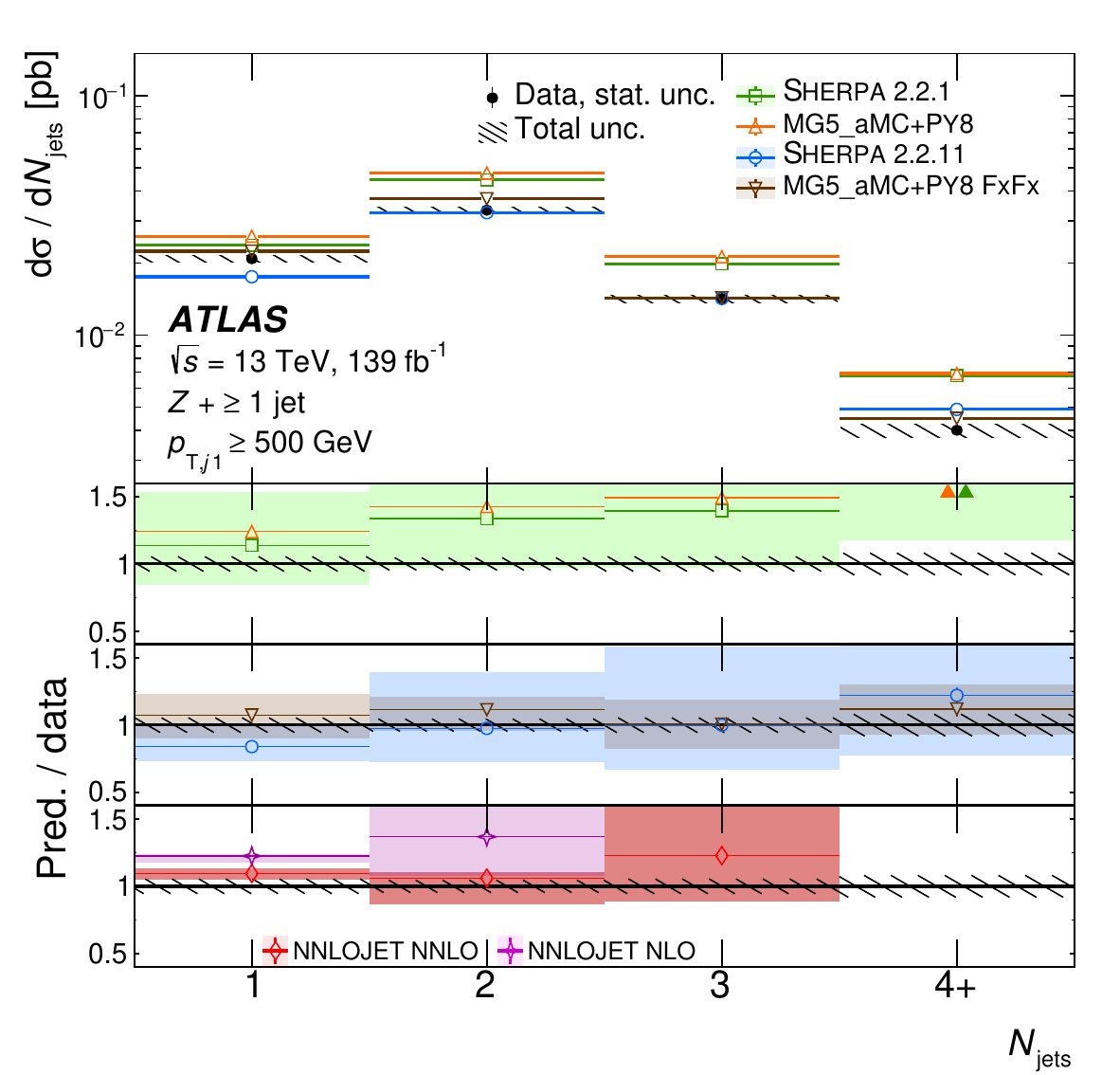}
}
\caption{
Differential cross section as a function of the jet multiplicity in the \inc\ region (left) and \hpt\ region (right). In each case, the last bin is inclusive in higher jet multiplicities. Further details are provided in the caption of Figure~\ref{fig:xs_withNLO_1}.
}
\label{fig:xs_withNLO_2}
\end{figure*}
 
The angular distance between the \ZBoson\ and the closest jet, \minDR, and the ratio of the \Zdash-boson transverse momentum to the closest-jet transverse momentum, \ptzj, provide a way to distinguish between the presence of collinear \Zdash-boson emission and back-to-back topologies. In the \hpt\ selection, the collinear region is sensitive to logarithmic enhancements in production proportional to $\alphas \ln^2(\ptlj/\mZ)$, whereas the back-to-back region receives non-negligible virtual EW corrections.
Figure~\ref{fig:xs_withNLO_3} shows the differential cross sections as a function of \minDR\ and \ptzj\ in the \hpt\ region. Both distributions show an accumulation of events at low and high values of these two quantities: \col\ events populate the figures at values $\minDR \leq 1.4$ and $\ptzj \leq 0.4$, while the \btb\ events are observed with $\minDR \approx \pi$ and $\ptzj \approx 1.0$.
The collinear events are expected to be dominated by diagrams corresponding to the EW radiation of a \ZBoson\ from one of the legs of a dijet event. Consequently, they are expected to correspond to the accumulation of events with low values of \ptzj. This hypothesis is validated in Figure~\ref{fig:xs_withNLO_4}, which shows the measurement of the \ptzj\ distribution for the subset of \col\ events defined by $\minDR \leq 1.4$ where only the accumulation of low-\ptzj\ events is observed. In contrast, the measurement of the \ptzj\ distribution for the \btb\ selection defined by $\minDR \geq 2.0$ is populated by events with $\ptzj \approx 1$. The jet multiplicity is also measured separately for the \col\ and \btb\ topologies as shown in Figure~\ref{fig:xs_withNLO_5}. It is found that the \col\ region is dominated by dijet events whereas the \btb\ region is dominated by \Zonejet\ events.
 
Figures~\ref{fig:xs_withNLO_3}--\ref{fig:xs_withNLO_5} show that while still marginally in agreement with data within modelling uncertainties of up to $50\%$, \Shtto\ central values increasingly overestimate the cross section with decreasing values of $\minDR$. A similar trend is observed for \MGflo.
The modelling is improved by the state-of-the-art generators \MGfnlo\ and \Shtte.
Good agreement of \MGfnlo\ and \Shtte\ with data even for very collinear events indicates that resummation of the additional logarithmic terms, e.g.\ via EW showers, is not needed at the present level of theoretical and experimental precision in the experimentally accessible kinematic regime. The \njnnlo\  prediction models the data distribution in both the \col\ and the \btb\ regions with high precision. The good performance in the \col\ phase space is remarkable, as this region contains a large fraction of events with at least three  jets, which are simulated only at LO in \njnnlo\ and not at all by \njnlo. The QCD-only calculation overestimates the cross section for  exact \btb\ events in the \hpt\ region, dominated by  high-\ptZ\ events, consistent with the pattern observed in Figure~\ref{fig:xs_withNLO_1}.
 
\begin{figure*}[t]
\centering
\subfloat{
\includegraphics[width=.50\textwidth]{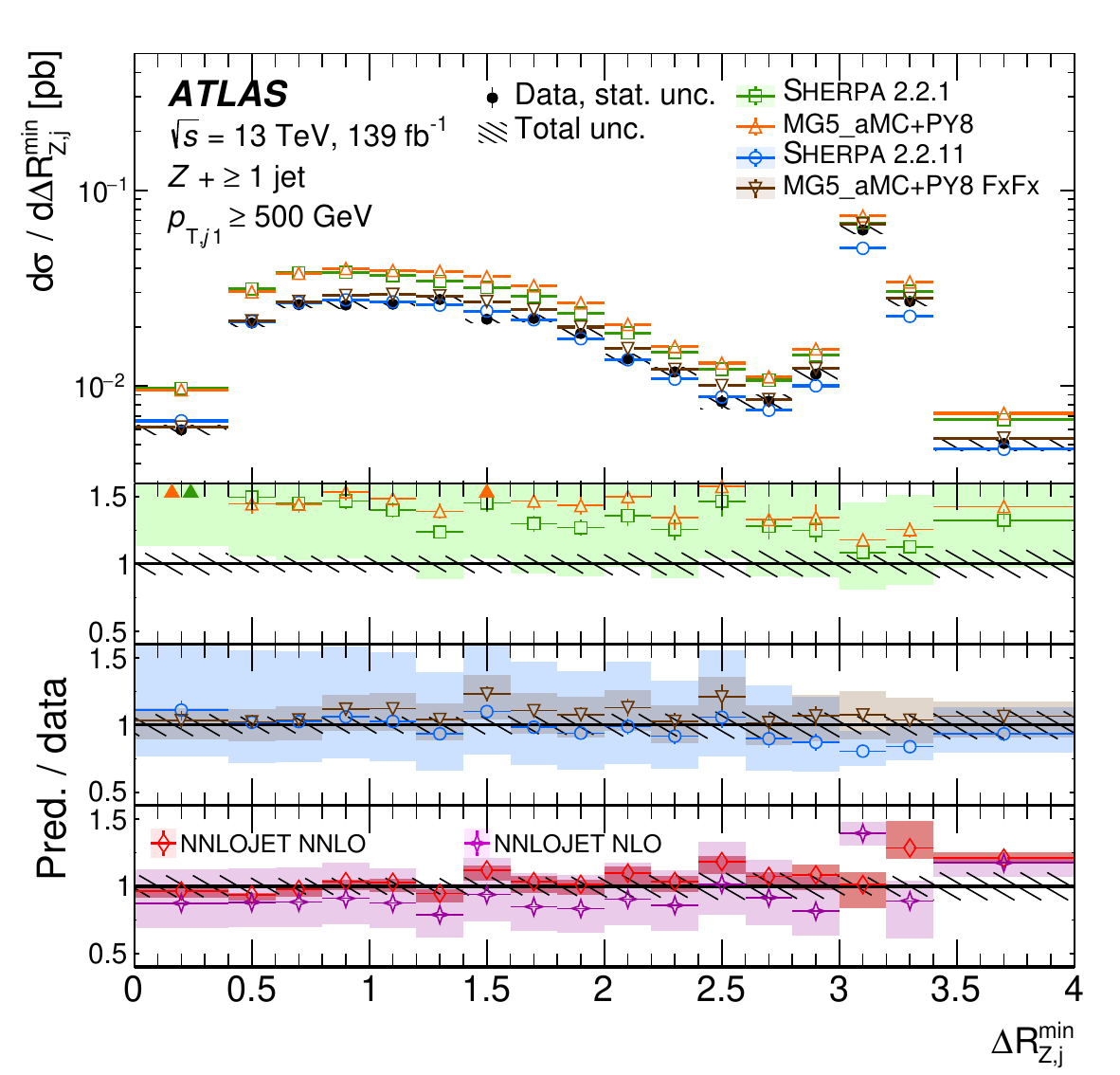}
}
\subfloat{
\includegraphics[width=.50\textwidth]{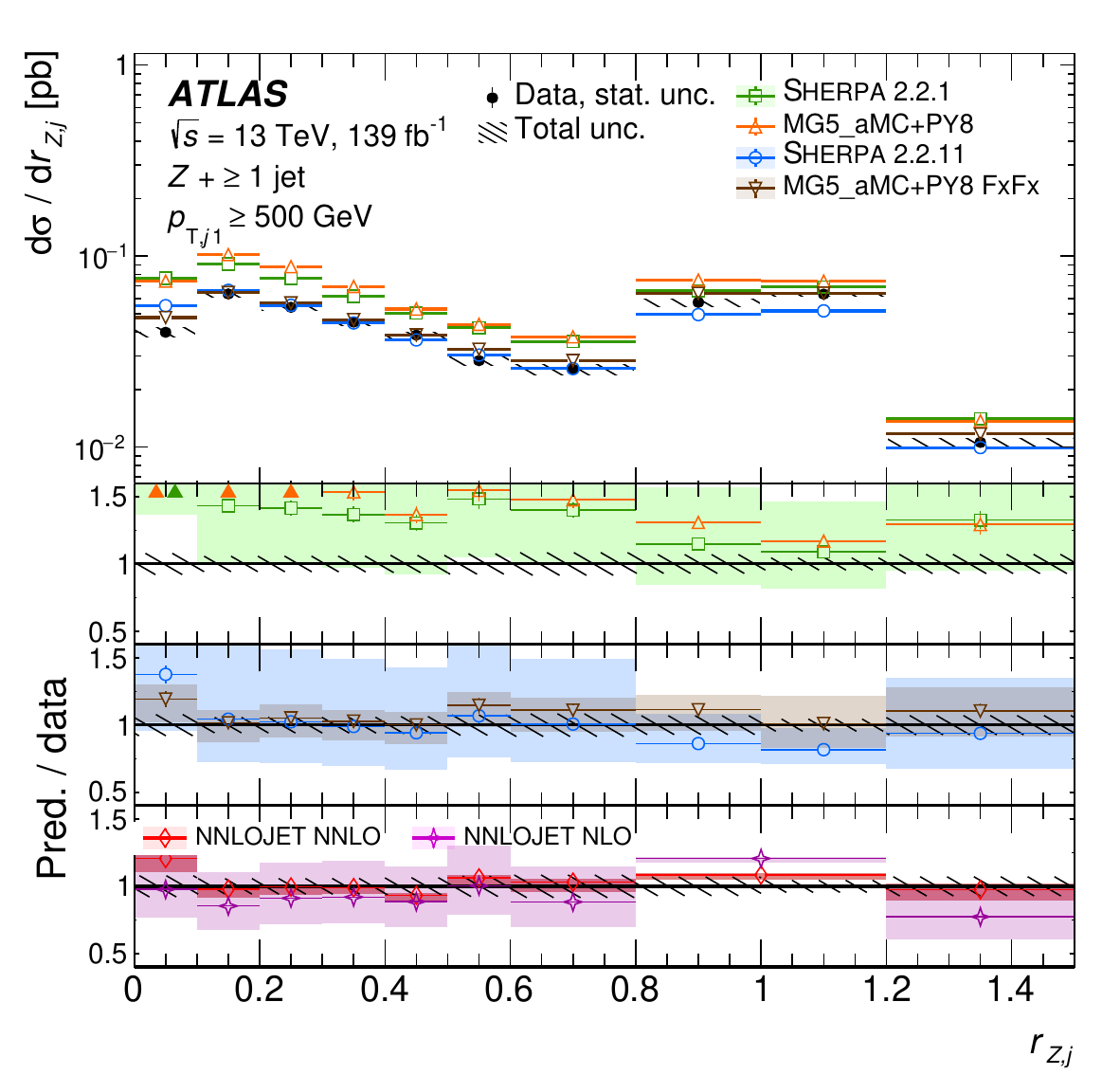}
}
\caption{
Differential cross section as a function of angular distance between the \ZBoson\ and the closest jet, \minDR\ (left) and of the ratio of \Zdash-boson to closest-jet transverse momenta \ptzj\ (right) in the \hpt\ region. For the \NNLOJET\ predictions, the bins around $\ptzj\ = 1$ are merged to be insensitive to the singularity in the fixed-order calculation. Further details are provided in the caption of Figure~\ref{fig:xs_withNLO_1}.
}
\label{fig:xs_withNLO_3}
\end{figure*}
 
\begin{figure*}[t]
\centering
\subfloat{
\includegraphics[width=.50\textwidth]{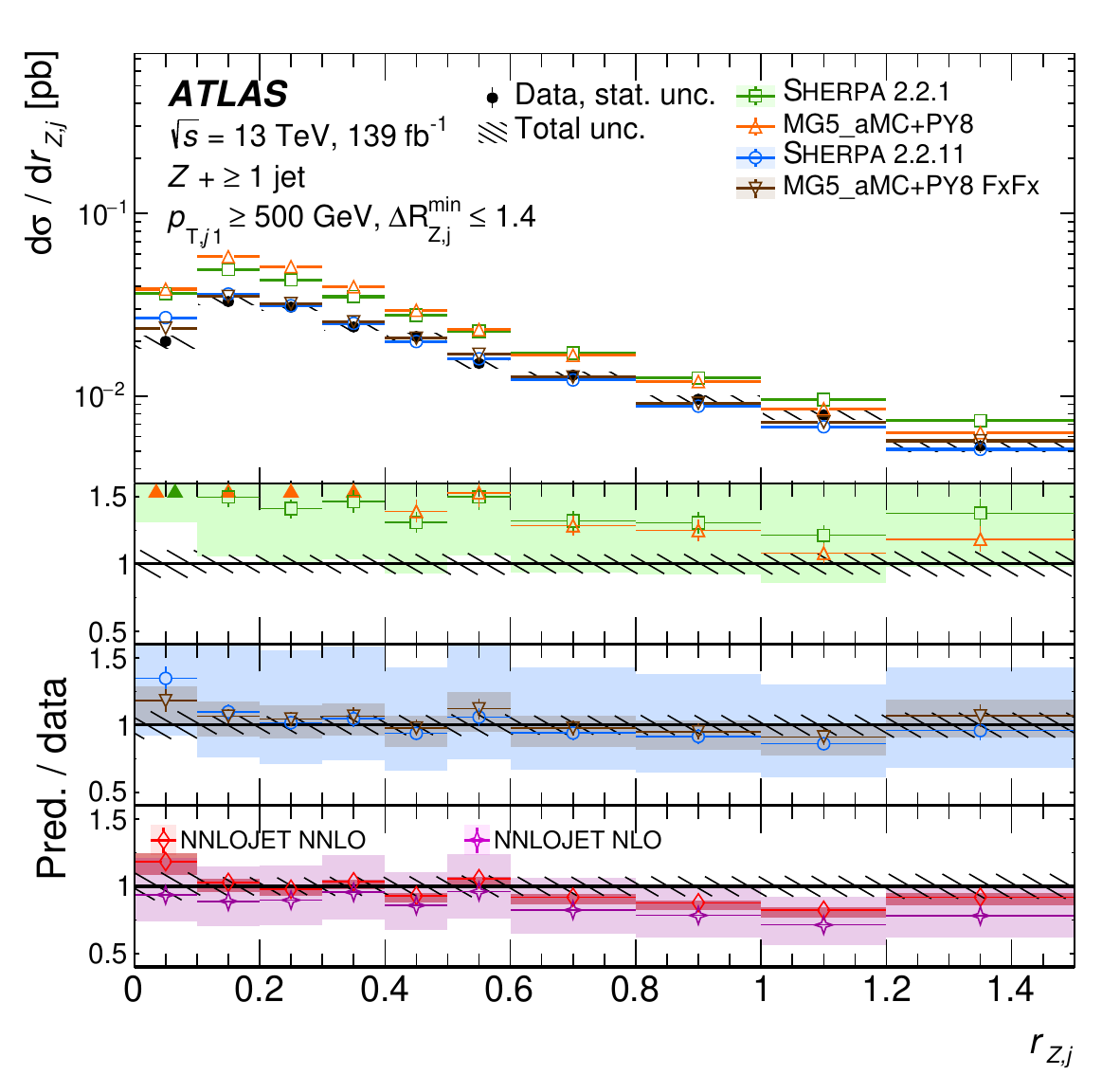}
}
\subfloat{
\includegraphics[width=.50\textwidth]{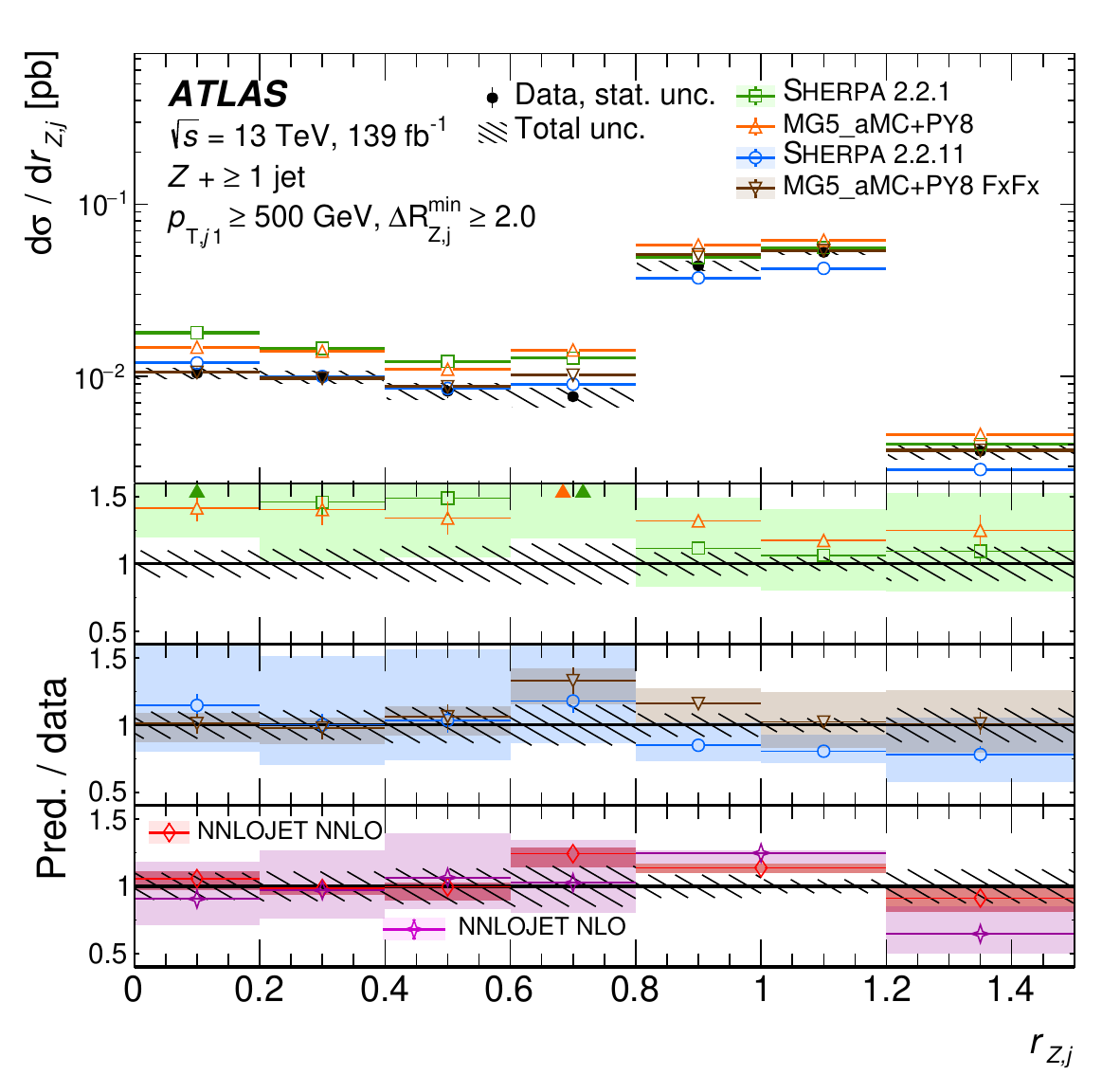}
}
\caption{
Differential cross section as a function of the ratio of \Zdash-boson to closest-jet transverse momenta $\ptzj$ in the \col\ $\DRZj \leq 1.4$ region (left) and in the \btb\ $\DRZj \geq 2.0$ region (right). For the \NNLOJET\ predictions, the bins around $\ptzj\ = 1$ are merged to be insensitive to the singularity in the fixed-order calculation. Further details are provided in the caption of Figure~\ref{fig:xs_withNLO_1}.
}
\label{fig:xs_withNLO_4}
\end{figure*}

\begin{figure*}[t]
\centering
\subfloat{
\includegraphics[width=.50\textwidth]{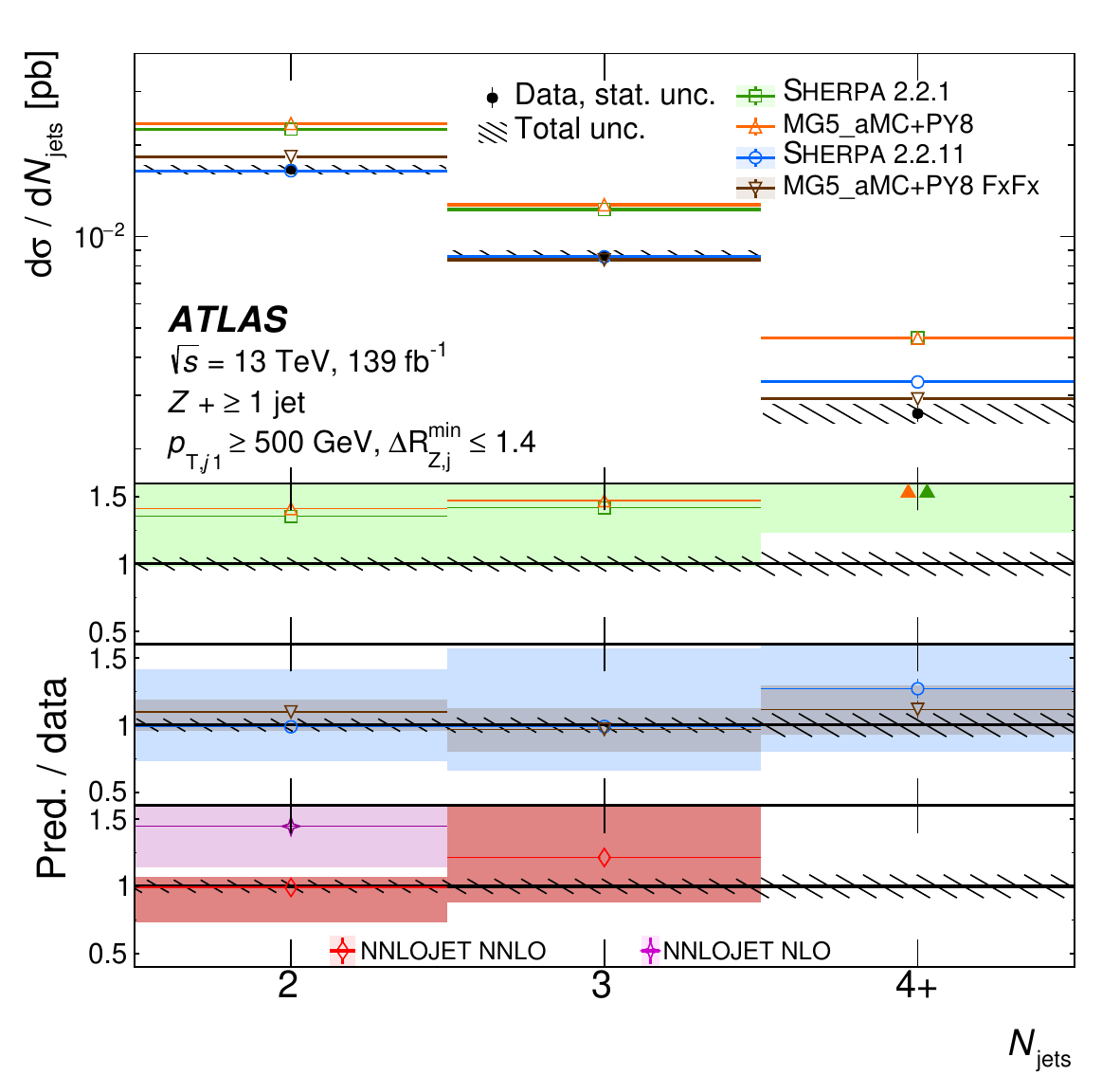}
}
\subfloat{
\includegraphics[width=.50\textwidth]{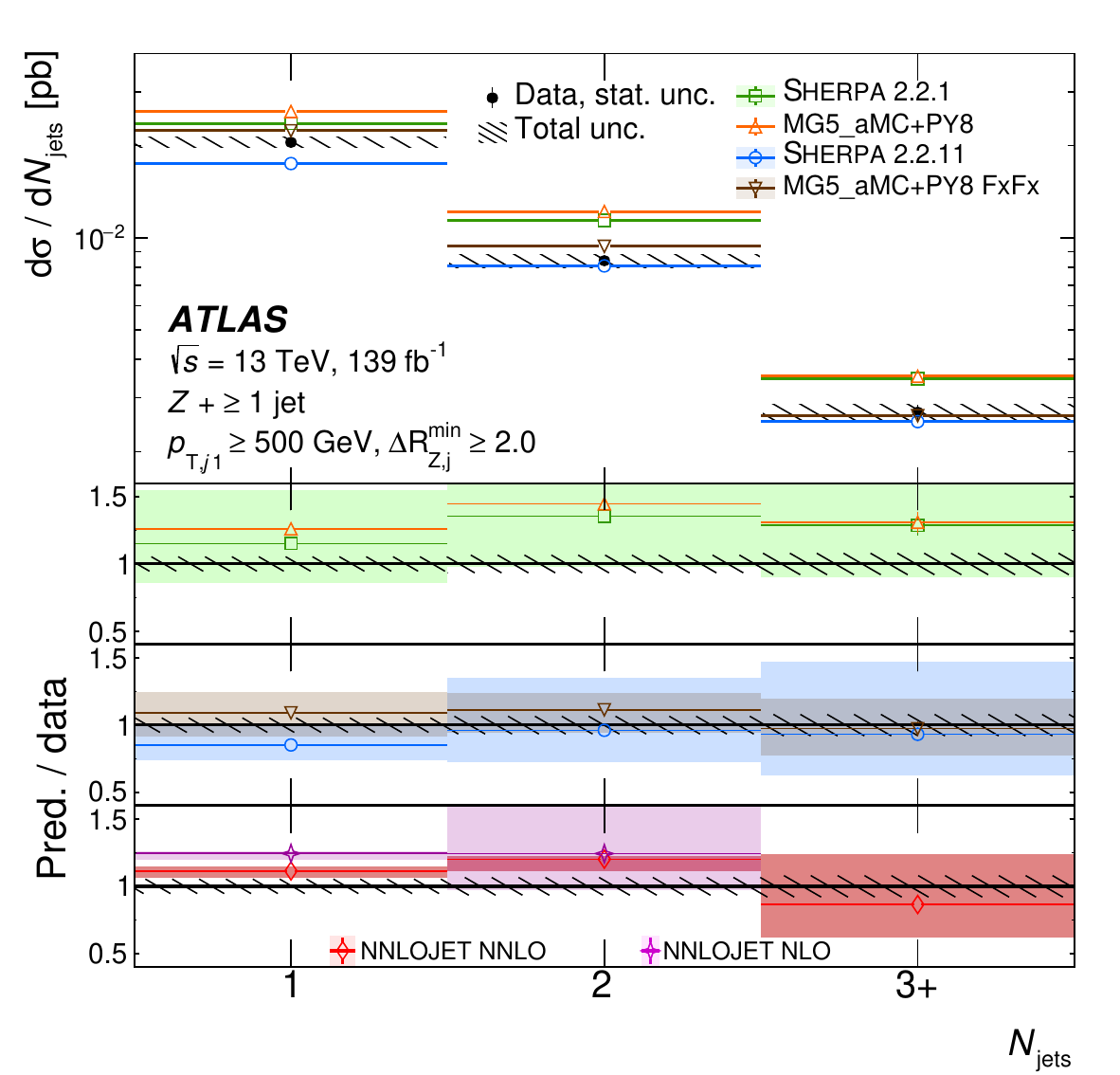}
}
\caption{Differential cross section as a function of the jet multiplicity in the \col\ $\DRZj \leq 1.4$ region (left) and in the \btb\ $\DRZj \geq 2.0$ region (right). In each case, the last bin is inclusive in higher jet multiplicities. In data, no events with exactly one jet are selected in the \col\ region. Further details are provided in the caption of Figure~\ref{fig:xs_withNLO_1}.
}
\label{fig:xs_withNLO_5}
\end{figure*}
 
An alternative way to select a high-energy phase space is by requiring a large value of \nht\ or \htj. The former is often used as a dynamical scale choice, whereas the latter is more suited to selecting a phase space similar to the \hpt\ region.
Figure~\ref{fig:xs_withNLO_6} shows the differential cross section as a function of \nht. The central values from \Shtto\ increasingly overestimate the cross section with increasing \nht, while still marginally agreeing with the data within modelling uncertainties of up to $50\%$. The prediction from \MGflo\ shows a similar trend.  In contrast, the state-of-the-art predictions from \Shtte, \MGfnlo\ and \njnnlo\ model the data well, the last with higher precision.
Figure~\ref{fig:xs_withNLO_7} shows the differential cross section as a function of \minDR\ and jet multiplicity in the \hht\ region. The \hht\ region probes high-energy events where the energy is typically shared by several jets. Compared to the \hpt\ region, the jet multiplicity distribution is shifted towards higher values, and the \btb\ peak in the \minDR\ distribution is suppressed relative to events where a jet is in closer proximity to the \ZBoson. As in the \hpt\ region, \Shtto\ is marginally consistent with the data within a large theoretical uncertainty, with the overestimation of the central values most pronounced in the \col\ region and for higher jet multiplicities,  with \MGflo\ showing a similar trend. Both \Shtte\ and \MGfnlo\ are consistent with the data, the latter at  higher precision. The \njnnlo\ prediction models this region well and with high precision, even though the fraction of events with at least three jets is larger than in the \hpt\ phase space.
 
\begin{figure*}[t]
\centering
\includegraphics[width=.50\textwidth]{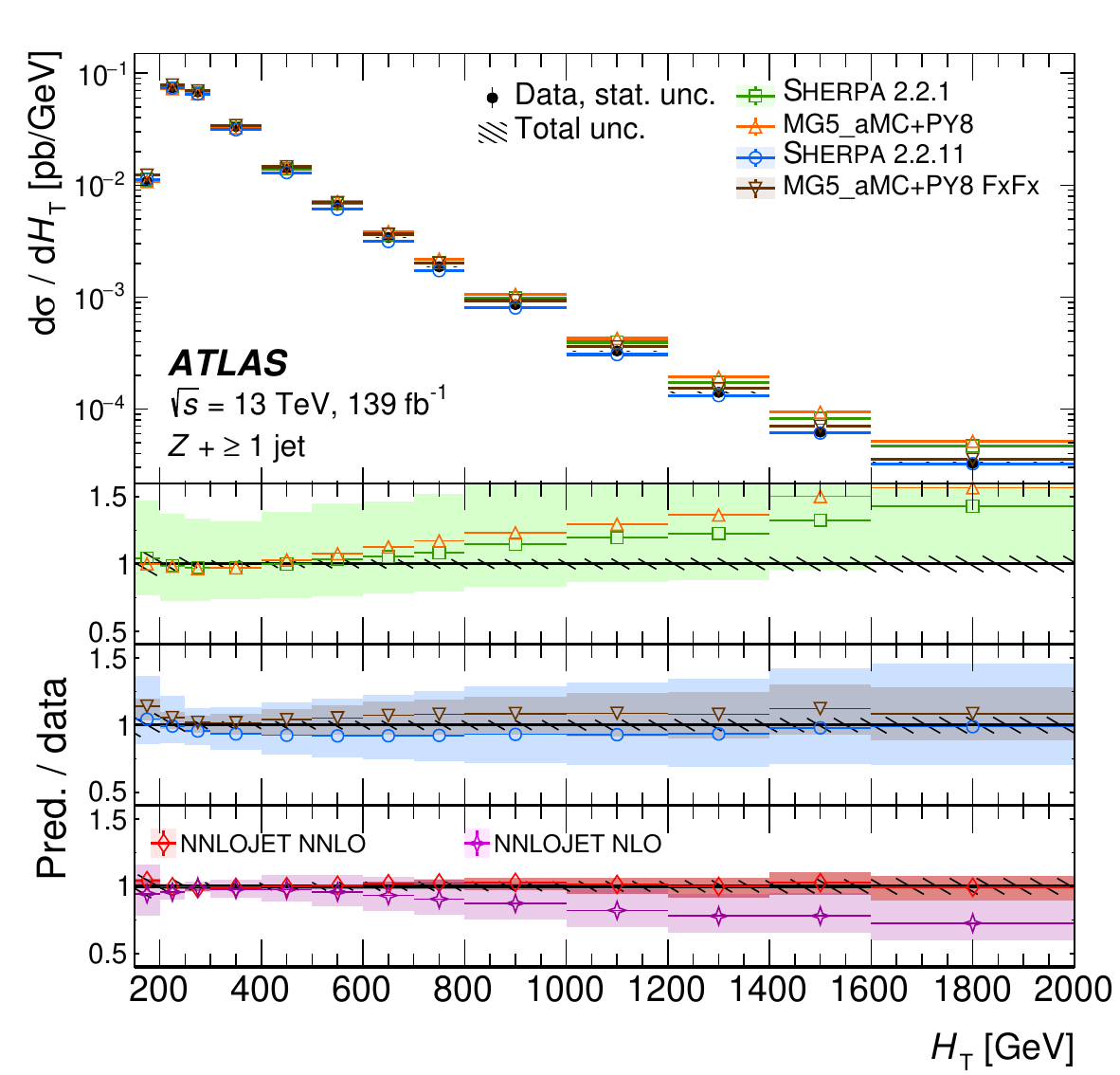}
\caption{
Differential cross section as a function of the transverse momenta scalar sum \nht\ in the \inc\ region. Further details are provided in the caption of Figure~\ref{fig:xs_withNLO_1}.
}
\label{fig:xs_withNLO_6}
\end{figure*}

\begin{figure*}[t]
\centering
\subfloat{
\includegraphics[width=.50\textwidth]{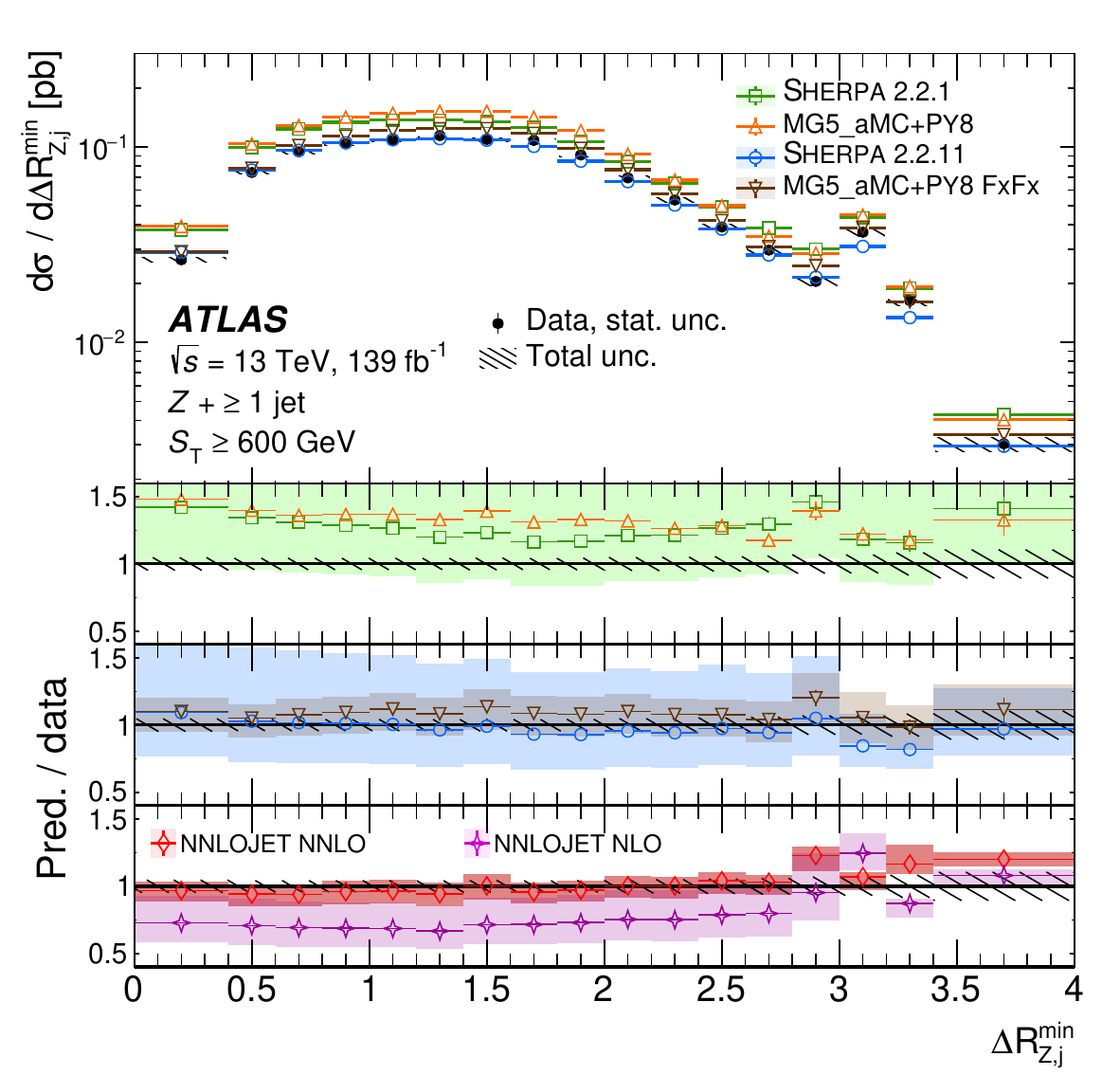}
}
\subfloat{
\includegraphics[width=.50\textwidth]{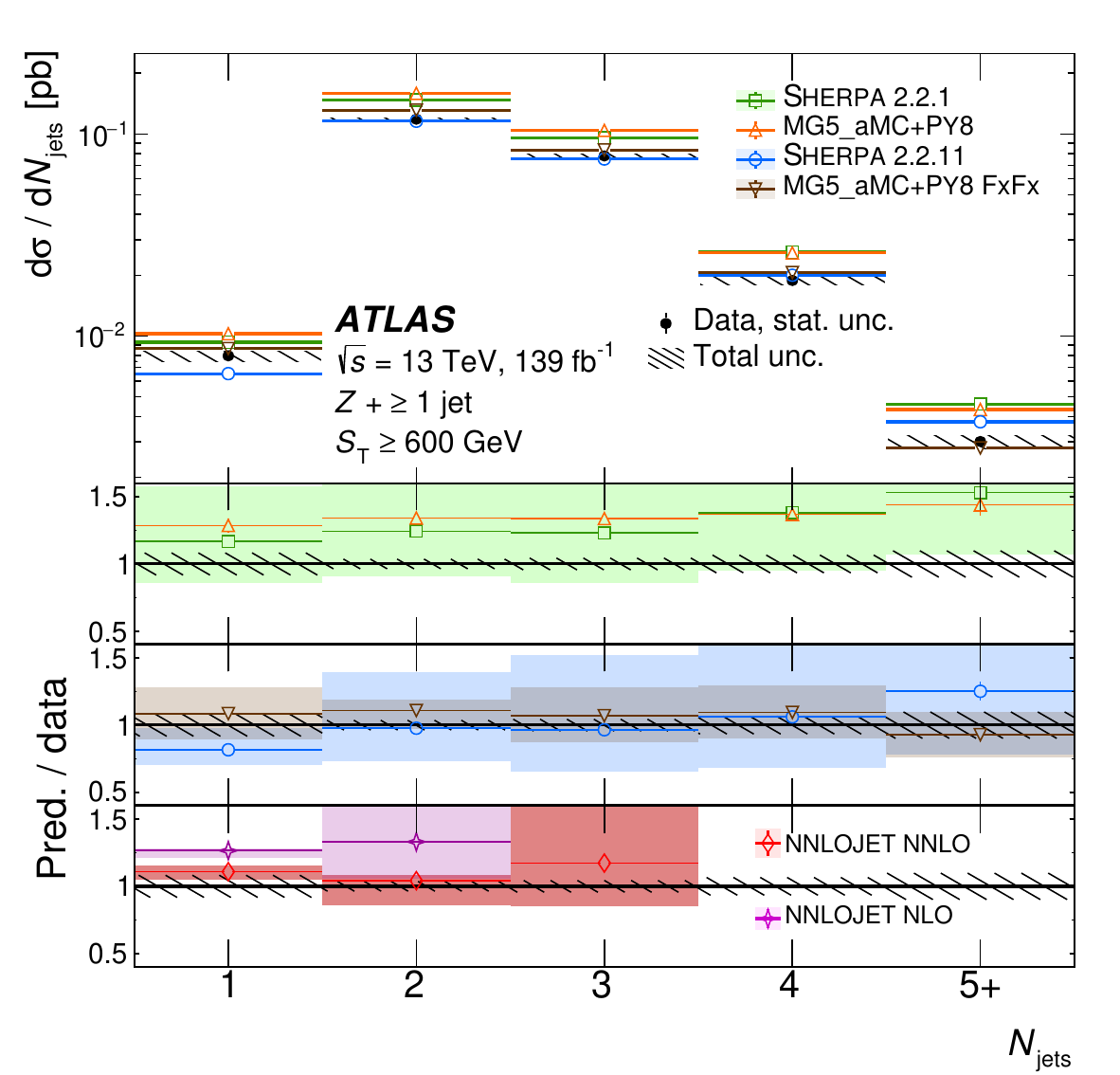}
}
\caption{
Differential cross section as a function of the angular distance \minDR\ between the \ZBoson\ and the closest jet (left) and of the jet multiplicity (right) in the \hht\ region. In the jet-multiplicity distribution, the last bin is inclusive in higher jet multiplicities. Further details are provided in the caption of Figure~\ref{fig:xs_withNLO_1}.
}
\label{fig:xs_withNLO_7}
\end{figure*}

\clearpage


\section{Conclusions}\label{sec:Conclusions}
 
This paper presents measurements of cross sections for a \ZBoson\ produced in association with high-transverse-momentum jets and decaying into a charged-lepton pair. The data were collected from $pp$ collisions at $\sqrt{s}= 13$~\TeV\ with the ATLAS detector at the LHC during 2015--2018 and correspond to an integrated luminosity of \Lumi. Measurements were performed on events that contain a \Zdash-boson candidate reconstructed from an $e^{+}e^{-}$ pair or a $\mu^{+}\mu^{-}$ pair in association with hadronic jets defined as jets having transverse momentum greater than or equal to $100~\GeV$. Primarily, only the QCD component of \Zjets\ production is measured, treating the EW $Zjj$ processes as a background. The paper focuses on selections with very high leading-jet \pT\ ($\ptlj \geq 500~\GeV$) and very high \htj\ ($\htj \geq 600~\GeV$), which are used to study two populations of events -- collinear events and back-to-back events -- with distinct patterns in distributions of \mDRZj, \ptzj, and jet multiplicity.
 
The data distributions were unfolded to the particle level and compared with state-of-the-art generator predictions and fixed-order calculations. Both \Shtto\ and \MGflo\ overestimate the cross sections for large values of  \ptlj, \nht\ and \htj. The predictions from \MGfnlo, with matrix elements for up to three partons at NLO, offer a significant improvement over \MGflo\ (which models up to four partons at LO) and in general match the data with high precision. Similarly, \Shtte, with the addition of a fifth parton at LO in the matrix element, the addition of NLO virtual EW corrections, and a different treatment of unordered histories in the parton shower, shows a significant improvement over \Shtto\ and agrees with the data. The NNLO calculations at fixed order from \NNLOJET\ describe the data cross sections at a very high level of precision, including in high-\pT regions where a sizeable fraction of the events have more than two jets. The calculation exhibits a harder \ptZ\ spectrum than the data in a region where larger negative EW corrections are expected.


\section*{Acknowledgements}


We thank CERN for the very successful operation of the LHC, as well as the
support staff from our institutions without whom ATLAS could not be
operated efficiently.
 
We acknowledge the support of
ANPCyT, Argentina;
YerPhI, Armenia;
ARC, Australia;
BMWFW and FWF, Austria;
ANAS, Azerbaijan;
CNPq and FAPESP, Brazil;
NSERC, NRC and CFI, Canada;
CERN;
ANID, Chile;
CAS, MOST and NSFC, China;
Minciencias, Colombia;
MEYS CR, Czech Republic;
DNRF and DNSRC, Denmark;
IN2P3-CNRS and CEA-DRF/IRFU, France;
SRNSFG, Georgia;
BMBF, HGF and MPG, Germany;
GSRI, Greece;
RGC and Hong Kong SAR, China;
ISF and Benoziyo Center, Israel;
INFN, Italy;
MEXT and JSPS, Japan;
CNRST, Morocco;
NWO, Netherlands;
RCN, Norway;
MEiN, Poland;
FCT, Portugal;
MNE/IFA, Romania;
MESTD, Serbia;
MSSR, Slovakia;
ARRS and MIZ\v{S}, Slovenia;
DSI/NRF, South Africa;
MICINN, Spain;
SRC and Wallenberg Foundation, Sweden;
SERI, SNSF and Cantons of Bern and Geneva, Switzerland;
MOST, Taiwan;
TENMAK, T\"urkiye;
STFC, United Kingdom;
DOE and NSF, United States of America.
In addition, individual groups and members have received support from
BCKDF, CANARIE, Compute Canada and CRC, Canada;
PRIMUS 21/SCI/017 and UNCE SCI/013, Czech Republic;
COST, ERC, ERDF, Horizon 2020 and Marie Sk{\l}odowska-Curie Actions, European Union;
Investissements d'Avenir Labex, Investissements d'Avenir Idex and ANR, France;
DFG and AvH Foundation, Germany;
Herakleitos, Thales and Aristeia programmes co-financed by EU-ESF and the Greek NSRF, Greece;
BSF-NSF and MINERVA, Israel;
Norwegian Financial Mechanism 2014-2021, Norway;
NCN and NAWA, Poland;
La Caixa Banking Foundation, CERCA Programme Generalitat de Catalunya and PROMETEO and GenT Programmes Generalitat Valenciana, Spain;
G\"{o}ran Gustafssons Stiftelse, Sweden;
The Royal Society and Leverhulme Trust, United Kingdom.
 
The crucial computing support from all WLCG partners is acknowledged gratefully, in particular from CERN, the ATLAS Tier-1 facilities at TRIUMF (Canada), NDGF (Denmark, Norway, Sweden), CC-IN2P3 (France), KIT/GridKA (Germany), INFN-CNAF (Italy), NL-T1 (Netherlands), PIC (Spain), ASGC (Taiwan), RAL (UK) and BNL (USA), the Tier-2 facilities worldwide and large non-WLCG resource providers. Major contributors of computing resources are listed in Ref.~\cite{ATL-SOFT-PUB-2021-003}.


\clearpage

\printbibliography

\clearpage
\section{HepData material}
 
The following tables will be included in the HEPData entry (\href{https://www.hepdata.net/record/114865}{link} to entry):
\begin{enumerate}
\item Differential cross section of \ptZ\ in the \inc\ region
\item Differential cross section of \ptlj\ in the \inc\ region
\item Differential cross section of jet multiplicity in the \inc\ region
\item Differential cross section of jet multiplicity in the \hpt\ region
\item Differential cross section of \minDR\ in the \hpt\ region
\item Differential cross section of \ptzj\ in the \hpt\ region
\item Differential cross section of \ptzj\ in the \col\ region
\item Differential cross section of \ptzj\ in the \btb\ region
\item Differential cross section of jet multiplicity in the \col\ region
\item Differential cross section of jet multiplicity in the \btb\ region
\item Differential cross section of \nht\ in the \inc\ region
\item Differential cross section of \minDR\ in the \hht\ region
\item Differential cross section of jet multiplicity in the \hht\ region
 
\item Relative bin-by-bin systematic uncertainties of \ptZ\ in the \inc\ region
\item Relative bin-by-bin systematic uncertainties of \ptlj\ in the \inc\ region
\item Relative bin-by-bin systematic uncertainties of jet multiplicity in the \inc\ region
\item Relative bin-by-bin systematic uncertainties of jet multiplicity in the \hpt\ region
\item Relative bin-by-bin systematic uncertainties of \minDR\ in the \hpt\ region
\item Relative bin-by-bin systematic uncertainties of \ptzj\ in the \hpt\ region
\item Relative bin-by-bin systematic uncertainties of \ptzj\ in the \col\ region
\item Relative bin-by-bin systematic uncertainties of \ptzj\ in the \btb\ region
\item Relative bin-by-bin systematic uncertainties of jet multiplicity in the \col\ region
\item Relative bin-by-bin systematic uncertainties of jet multiplicity in the \btb\ region
\item Relative bin-by-bin systematic uncertainties of \nht\ in the \inc\ region
\item Relative bin-by-bin systematic uncertainties of \minDR\ in the \hht\ region
\item Relative bin-by-bin systematic uncertainties of jet multiplicity in the \hht\ region
 
\item Bin-by-bin Born to dressed level leptons correction scale factor for \ptZ\ in the \inc\ region
\item Bin-by-bin Born to dressed level leptons correction scale factor for \ptlj\ in the \inc\ region
\item Bin-by-bin Born to dressed level leptons correction scale factor for the jet multiplicity in the \inc\ region
\item Bin-by-bin Born to dressed level leptons correction scale factor for the jet multiplicity in the \hpt\ region
\item Bin-by-bin Born to dressed level leptons correction scale factor for \minDR\ in the \hpt\ region
\item Bin-by-bin Born to dressed level leptons correction scale factor for \ptzj\ in the \hpt\ region
\item Bin-by-bin Born to dressed level leptons correction scale factor for \ptzj\ in the \col\ region
\item Bin-by-bin Born to dressed level leptons correction scale factor for \ptzj\ in the \btb\ region
\item Bin-by-bin Born to dressed level leptons correction scale factor for the jet multiplicity in the \col\ region
\item Bin-by-bin Born to dressed level leptons correction scale factor for the jet multiplicity in the \btb\ region
\item Bin-by-bin Born to dressed level leptons correction scale factor for \nht\ in the \inc\ region
\item Bin-by-bin Born to dressed level leptons correction scale factor for \minDR\ in the \hht\ region
\item Bin-by-bin Born to dressed level leptons correction scale factor for the jet multiplicity in the \hht\ region
 
\item Bin-by-bin overlap removal correction scale factor for \ptZ\ in the \inc\ region
\item Bin-by-bin overlap removal correction scale factor for \ptlj\ in the \inc\ region
\item Bin-by-bin overlap removal correction scale factor for the jet multiplicity in the \inc\ region
\item Bin-by-bin overlap removal correction scale factor for the jet multiplicity in the \hpt\ region
\item Bin-by-bin overlap removal correction scale factor for \minDR\ in the \hpt\ region
\item Bin-by-bin overlap removal correction scale factor for \ptzj\ in the \hpt\ region
\item Bin-by-bin overlap removal correction scale factor for \ptzj\ in the \col\ region
\item Bin-by-bin overlap removal correction scale factor for \ptzj\ in the \btb\ region
\item Bin-by-bin overlap removal correction scale factor for the jet multiplicity in the \col\ region
\item Bin-by-bin overlap removal correction scale factor for the jet multiplicity in the \btb\ region
\item Bin-by-bin overlap removal correction scale factor for \nht\ in the \inc\ region
\item Bin-by-bin overlap removal correction scale factor for \minDR\ in the \hht\ region
\item Bin-by-bin overlap removal correction scale factor for the jet multiplicity in the \hht\ region
 
\item Differential cross section of \ptZ\ in the \inc\ region, where the EW Zjj contribution is not subtracted as background
\item Differential cross section of \ptlj\ in the \inc\ region, where the EW Zjj contribution is not subtracted as background
\item Differential cross section of jet multiplicity in the \inc\ region, where the EW Zjj contribution is not subtracted as background
\item Differential cross section of jet multiplicity in the \hpt\ region, where the EW Zjj contribution is not subtracted as background
\item Differential cross section of \minDR\ in the \hpt\ region, where the EW Zjj contribution is not subtracted as background
\item Differential cross section of \ptzj\ in the \hpt\ region, where the EW Zjj contribution is not subtracted as background
\item Differential cross section of \ptzj\ in the \col\ region, where the EW Zjj contribution is not subtracted as background
\item Differential cross section of \ptzj\ in the \btb\ region, where the EW Zjj contribution is not subtracted as background
\item Differential cross section of jet multiplicity in the \col\ region, where the EW Zjj contribution is not subtracted as background
\item Differential cross section of jet multiplicity in the \btb\ region, where the EW Zjj contribution is not subtracted as background
\item Differential cross section of \nht\ in the \inc\ region, where the EW Zjj contribution is not subtracted as background
\item Differential cross section of \minDR\ in the \hht\ region, where the EW Zjj contribution is not subtracted as background
\item Differential cross section of jet multiplicity in the \hht\ region, where the EW Zjj contribution is not subtracted as background
 
\item Relative bin-by-bin systematic uncertainties of \ptZ\ in the \inc\ region, where the EW Zjj contribution is not subtracted as background
\item Relative bin-by-bin systematic uncertainties of \ptlj\ in the \inc\ region, where the EW Zjj contribution is not subtracted as background
\item Relative bin-by-bin systematic uncertainties of jet multiplicity in the \inc\ region, where the EW Zjj contribution is not subtracted as background
\item Relative bin-by-bin systematic uncertainties of jet multiplicity in the \hpt\ region, where the EW Zjj contribution is not subtracted as background
\item Relative bin-by-bin systematic uncertainties of \minDR\ in the \hpt\ region, where the EW Zjj contribution is not subtracted as background
\item Relative bin-by-bin systematic uncertainties of \ptzj\ in the \hpt\ region, where the EW Zjj contribution is not subtracted as background
\item Relative bin-by-bin systematic uncertainties of \ptzj\ in the \col\ region, where the EW Zjj contribution is not subtracted as background
\item Relative bin-by-bin systematic uncertainties of \ptzj\ in the \btb\ region, where the EW Zjj contribution is not subtracted as background
\item Relative bin-by-bin systematic uncertainties of jet multiplicity in the \col\ region, where the EW Zjj contribution is not subtracted as background
\item Relative bin-by-bin systematic uncertainties of jet multiplicity in the \btb\ region, where the EW Zjj contribution is not subtracted as background
\item Relative bin-by-bin systematic uncertainties of \nht\ in the \inc\ region, where the EW Zjj contribution is not subtracted as background
\item Relative bin-by-bin systematic uncertainties of \minDR\ in the \hht\ region, where the EW Zjj contribution is not subtracted as background
\item Relative bin-by-bin systematic uncertainties of jet multiplicity in the \hht\ region, where the EW Zjj contribution is not subtracted as background

\end{enumerate}

\clearpage

\clearpage
 
\input{atlas_authlist}

\end{document}

%% file: atlas_authlist.tex
 
\begin{flushleft}
\hypersetup{urlcolor=black}
{\Large The ATLAS Collaboration}

\bigskip

\AtlasOrcid[0000-0002-6665-4934]{G.~Aad}$^\textrm{\scriptsize 101}$,
\AtlasOrcid[0000-0002-5888-2734]{B.~Abbott}$^\textrm{\scriptsize 119}$,
\AtlasOrcid[0000-0002-7248-3203]{D.C.~Abbott}$^\textrm{\scriptsize 102}$,
\AtlasOrcid[0000-0002-1002-1652]{K.~Abeling}$^\textrm{\scriptsize 55}$,
\AtlasOrcid[0000-0002-8496-9294]{S.H.~Abidi}$^\textrm{\scriptsize 29}$,
\AtlasOrcid[0000-0002-9987-2292]{A.~Aboulhorma}$^\textrm{\scriptsize 35e}$,
\AtlasOrcid[0000-0001-5329-6640]{H.~Abramowicz}$^\textrm{\scriptsize 150}$,
\AtlasOrcid[0000-0002-1599-2896]{H.~Abreu}$^\textrm{\scriptsize 149}$,
\AtlasOrcid[0000-0003-0403-3697]{Y.~Abulaiti}$^\textrm{\scriptsize 116}$,
\AtlasOrcid[0000-0003-0762-7204]{A.C.~Abusleme~Hoffman}$^\textrm{\scriptsize 136a}$
% ...
\bigskip
\\

% ...
$^{29}$Physics Department, Brookhaven National Laboratory, Upton NY; United States of America.\\
% ...
$^{35}$$^{(a)}$Facult\'e des Sciences Ain Chock, R\'eseau Universitaire de Physique des Hautes Energies - Universit\'e Hassan II, Casablanca;$^{(b)}$Facult\'{e} des Sciences, Universit\'{e} Ibn-Tofail, K\'{e}nitra;$^{(c)}$Facult\'e des Sciences Semlalia, Universit\'e Cadi Ayyad, LPHEA-Marrakech;$^{(d)}$LPMR, Facult\'e des Sciences, Universit\'e Mohamed Premier, Oujda;$^{(e)}$Facult\'e des sciences, Universit\'e Mohammed V, Rabat;$^{(f)}$Institute of Applied Physics, Mohammed VI Polytechnic University, Ben Guerir; Morocco.\\
% ...
$^{55}$II. Physikalisches Institut, Georg-August-Universit\"{a}t G\"ottingen, G\"ottingen; Germany.\\
% ...
$^{101}$CPPM, Aix-Marseille Universit\'e, CNRS/IN2P3, Marseille; France.\\
$^{102}$Department of Physics, University of Massachusetts, Amherst MA; United States of America.\\
% ...
$^{116}$Department of Physics, New York University, New York NY; United States of America.\\
% ...
$^{119}$Homer L. Dodge Department of Physics and Astronomy, University of Oklahoma, Norman OK; United States of America.\\
% ...
$^{136}$$^{(a)}$Departamento de F\'isica, Pontificia Universidad Cat\'olica de Chile, Santiago;$^{(b)}$Millennium Institute for Subatomic physics at high energy frontier (SAPHIR), Santiago;$^{(c)}$Instituto de Investigaci\'on Multidisciplinario en Ciencia y Tecnolog\'ia, y Departamento de F\'isica, Universidad de La Serena;$^{(d)}$Universidad Andres Bello, Department of Physics, Santiago;$^{(e)}$Instituto de Alta Investigaci\'on, Universidad de Tarapac\'a, Arica;$^{(f)}$Departamento de F\'isica, Universidad T\'ecnica Federico Santa Mar\'ia, Valpara\'iso; Chile.\\
% ...
$^{149}$Department of Physics, Technion, Israel Institute of Technology, Haifa; Israel.\\
$^{150}$Raymond and Beverly Sackler School of Physics and Astronomy, Tel Aviv University, Tel Aviv; Israel.\\
% ...
$^{a}$ Also Affiliated with an institute covered by a cooperation agreement with CERN.\\
$^{b}$ Also at Borough of Manhattan Community College, City University of New York, New York NY; United States of America.\\
$^{c}$ Also at Bruno Kessler Foundation, Trento; Italy.\\
$^{d}$ Also at Center for High Energy Physics, Peking University; China.\\
$^{e}$ Also at Centro Studi e Ricerche Enrico Fermi; Italy.\\
$^{f}$ Also at CERN, Geneva; Switzerland.\\
% ...

\end{flushleft}
